\documentclass[aps,pra,reprint,superscriptaddress,footinbib,longbibliography]{revtex4-2}

\usepackage{amssymb}
\usepackage{amsmath}
\usepackage{amsfonts}
\usepackage{xcolor}
\usepackage{graphicx}
\usepackage{bm}
\usepackage{physics}
\usepackage{stackengine}
\usepackage{enumitem}
\usepackage{mathtools}
\usepackage[normalem]{ulem}
\usepackage{verbatim}

\usepackage[unicode]{hyperref}
\hypersetup{
   unicode=true,          
   plainpages=false,
   colorlinks=true,       
   citecolor=blue,        
}

\allowdisplaybreaks

\newcommand*{\vek}[1]{{\ensuremath{\bm{\mathrm{#1}}}}}
\newcommand\barb[1]{\stackunder[1.2pt]{\ensuremath{#1}}{\rule{.8ex}{.075ex}}}
\newcommand\bbarb[1]{\barb{\barb{#1}}}

\newcommand{\Langle}{\big\langle}
\newcommand{\Rangle}{\big\rangle}
\newcommand{\rr}{\mathbf{r}}

\newcommand{\Dproj}{\bbarb{\mathcal{P}}}
\newcommand{\Cproj}{\bbarb{\mathcal{Q}}}

\begin{document}

\title{Odd-frequency superfluidity from a
particle-number-conserving perspective}

\author{K. Thompson}
\affiliation{MacDiarmid Institute, School of Chemical and
Physical Sciences, Victoria University of Wellington, PO
Box 600, Wellington 6140, New Zealand}
\affiliation{Cavendish Laboratory, JJ Thomson Avenue,
Cambridge CB3 0HE, United Kingdom}

\author{U. Z\"ulicke}
\email{uli.zuelicke@vuw.ac.nz}
\affiliation{MacDiarmid Institute, School of Chemical and
Physical Sciences, Victoria University of Wellington, PO
Box 600, Wellington 6140, New Zealand}
\affiliation{Dodd-Walls Centre for Photonic and Quantum
Technologies, School of Chemical and Physical Sciences,
Victoria University of Wellington, PO Box 600, Wellington
6140, New Zealand}

\author{J. Schmalian}
\affiliation{Institute for Theory of Condensed
Matter, Karlsruhe Institute of Technology, 76131
Karlsruhe,Germany}
\affiliation{Institute for Quantum Materials and
Technologies, Karlsruhe Institute of Technology, 76126
Karlsruhe, Germany}

\author{M. Governale}
\affiliation{MacDiarmid Institute, School of Chemical and
Physical Sciences, Victoria University of Wellington, PO
Box 600, Wellington 6140, New Zealand}

\author{J. Brand}
\email{j.brand@massey.ac.nz}
\affiliation{Dodd-Walls Centre for Photonic and Quantum
Technologies, Centre for Theoretical Chemistry and
Physics, New Zealand Institute for Advanced Study,
Massey University, Private Bag 102904, North Shore,
Auckland 0745, New Zealand}

\date{\today}

\begin{abstract}
We investigate odd-in-time---or
\emph{odd-frequency}---pairing of fermions in
equilibrium systems within the
particle-number-conserving framework of Penrose,
Onsager and Yang, where superfluid order is defined by
macroscopic eigenvalues of reduced density matrices. We
show that odd-frequency pair correlations are synonymous
with even fermion-exchange symmetry in a time-dependent
correlation function that generalises the two-body
reduced density matrix. Macroscopic
even-under-fermion-exchange pairing is found to emerge
from conventional Penrose-Onsager-Yang condensation in
two-body or higher-order reduced density matrices
through the symmetry-mixing properties of the
Hamiltonian. We identify and characterise a
\emph{transformer} matrix responsible for producing
macroscopic even fermion-exchange correlations that
coexist with a conventional Cooper-pair condensate,
while a \emph{generator} matrix is shown to be
responsible for creating macroscopic even
fermion-exchange correlations from hidden orders such as
a multi-particle condensate. The transformer scenario is
illustrated using the spin-balanced
\textit{s}-wave superfluid with Zeeman splitting as an
example. The generator scenario is demonstrated by the
composite-boson condensate arising for itinerant
electrons coupled to magnetic excitations. Structural
analysis of the transformer and generator matrices is
shown to provide general conditions for odd-frequency
pairing order to arise in a given system. 
Our formalism facilitates a fully general derivation of
the Meissner effect for odd-frequency superconductors
that holds also beyond the regime of validity for
mean-field theory.

\end{abstract}

\maketitle

\section{Introduction}\label{sec:intro}

Superconductivity is a striking phenomenon whose
macroscopic consequences include zero electric
resistivity, the Meissner and Josephson effects, and
magnetic-flux quantization~\cite{Ketterson1999}. The
mechanism of conventional superconductivity is well
understood~\cite{Bardeen1957}: bosonic Cooper pairs with
\textit{s}-wave symmetry are formed by phonon-mediated
attraction between electrons, and these bosons condense
at low temperatures. Unconventional superconductors,
which do not conform to the same pattern, have been the
subject of active research for decades~\cite{Sigrist1991,
Mineev1999}. Known unconventional superconductors
involve Cooper pairs with different orbital symmetry
(e.g., \textit{p}-wave or \textit{d}-wave), or
alternative pairing mechanisms like repulsive
electron-electron interactions or spin fluctuations.
Odd-frequency superconductivity is an exotic
hypothesized form of unconventional
superconductivity~\cite{Linder2019}. Going back to an
attempt by Berezinskii to explain superfluidity in
${}^3$He~\cite{Berezinskii1974}, odd-frequency
superconductivity is based on the mathematical
possibility that natural systems might be described by a
finite anomalous pair correlation
function~\cite{Gorkov1958,Bruus2004}
\begin{align}\label{eq:AnomCorr}
F_{\vek{i}\vek{j}}(t_1,t_2) = \langle T\,
c^\dagger_\vek{i}(t_1)\, c^\dagger_\vek{j}(t_2) \rangle
\end{align}
with odd symmetry under exchange of the time arguments,
which also implies an even symmetry under exchange of
the fermion indices. This is in contrast to the standard
theories of superconductivity (both conventional and
unconventional), which are based on equal-time pair
correlation functions with odd pair-exchange symmetry.
Here $\vek{i}$ and $\vek{j}$ denote sets of indices used
for labeling single-fermion quantum numbers, and $T$ is
the time-ordering operator. The name ``odd-frequency''
refers to the fact that an odd symmetry under exchange
of the time arguments [i.e., $F_{\vek{i}\vek{j}}(t_1,
t_2)$ satisfying $F_{\vek{i}\vek{j}}(t_1,t_2) =
-F_{\vek{i}\vek{j}}(t_2,t_1)$] implies an odd symmetry
in the frequency domain. 

Even though some theoretical models suggest that
odd-frequency superconductivity may exist in bulk
materials and in the absence of conventional
(even-frequency) superconductivity
\cite{Berezinskii1974,Kirkpatrick1991,Balatsky1992,
Abrahams1993,Emery1992,Balatsky1993,Coleman1994,
Coleman1995,Balatsky1994,Schrieffer1994,Abrahams1995,
Dahal2009,Fuseya2003,Kusunose2011a}, and despite an
intense search for such phases, no conclusive evidence
has been found to date~\cite{Linder2019}. In a separate
scenario, odd-frequency pair correlations have been
proposed to occur in the presence of a conventional
even-frequency Cooper-pair condensate when translational
and/or spin-rotational symmetries are
broken~\cite{Bergeret2005,Eschrig2007,Tanaka2007,
Kusunose2012,Sothmann2014,Black-Schaffer2013,
Asano2015,Triola2020,Chakraborty2022inter}. This
alternative is supported by indirect experimental
evidence for odd-frequency correlations in
heterostructures and near defects, through spectroscopic
measurements and observations of the density of
states~\cite{DiBernardo2015,Pal2017,Perrin2020}. Some
theoretical understanding of the apparent absence of bulk
odd-frequency superconductors has been obtained through
a discussion about the thermodynamic stability of
odd-frequency superconducting states \cite{Heid1995,
Solenov2009,Kusunose2011,Fominov2015}, and the recent
development of \emph{no-go} theorems within the
framework of Eliashberg theory~\cite{Schrodi2021,
Pimenov2022,Langmann2022}. Thus, whether odd-frequency
superconductivity is realized in nature still remains an
open question.

Existing theories of odd-frequency superconductivity are
deeply rooted in the formalism of anomalous Greens
functions~\cite{Bruus2004}, which violates
particle-number conservation for electrons. This is
potentially problematic, as the electron number is
fundamentally expected to be
conserved~\cite{Leggett2006}. While number-nonconserving
theories of superconductivity and superfluidity have
been very successful in describing and predicting many
important phenomena~\cite{Anderson1969}, they are at
odds with nature~\cite{Wightman1995} and thus not fully
satisfactory. Moreover, with the recent development of
ultracold-atom experiments in optical lattices and
microtraps, it has become possible to study
superfluidity in systems where the particle number is
fixed~\cite{Bloch2008a,Bloch2012,Altman2021,Li2024} and
small, even down to single digits \cite{Holten2022},
such that the consequences of number fluctuations---or
their absence---matter.  It is thus important to ask
whether odd-frequency superconductivity can be described
using a particle-number conserving formalism, and what
this would look like. It is the purpose of this paper to
take the first step in answering this question. To this
end, we develop a number-conserving formalism of
superfluid and superconducting states based on a
time-dependent two-particle correlation function. This
work generalises the Penrose-Onsager-Yang criterion for
fermionic superfluidity~\cite{Penrose1956,Yang1962},
which is based on the presence of a macroscopic
eigenvalue of the two-particle reduced density matrix,
i.e., one that scales linearly with the particle number
$N$~\footnote{We use the term ``macroscopic'' in the
sense that a macroscopic quantity is expected to grow
proportionally to the number of fermions $N$ when the
system size is increased. For a mathematically precise
definition, see Appendix~\ref{app:genEvT2bCMEv}. The
actual value of $N$ could be of the order of Avogadro's
number, e.g., for solid-state systems, but it is
typically much smaller (and thus not truly macroscopic
in the thermodynamic sense) in ultracold-atom
experiments.}. Our extension includes a relative-time
dependence of pairing between two particles. It thus
provides a natural framework for the study of
odd-frequency superfluidity, while it reduces to the
usual criterion in the case of even-frequency
superfluidity~\footnote{There exist also other
particle-conserving theories for conventional
superfluidity~\cite{Richardson1964,vonDelft2000,
Janssen1981,Koizumi2021}. Here we choose to follow the
Penrose-Onsager-Yang approach because of its
versatility~\cite{Leggett2006}.}. 
The relative time in this formalism refers to a
time delay for probing the fermion-pair correlations and
not the time scale of nonequilibrium dynamics of the
host system.

A particle-number-conserving description
also pertains to the important concept of
off-diagonal long-range order~\cite{Yang1962}, which not
only demonstrates that a superconductor is a state of
macroscopic quantum coherence, but also allows to draw
conclusions about phenomena such as flux quantization,
the Meissner effect, and related phenomena, even if
conventional mean-field approaches do not apply. This
aspect is  particularly important for states with exotic
order, such as odd-frequency pairing states. In
fact, an important conclusion drawn from
our formalism is the existence of a diamagnetic Meissner
effect for (even- and) odd-frequency superconductors.
Generalizing well-known arguments for conventional
superconductors~\cite{Sewell1990}, we find that the
presence of off-diagonal long-range order at any value
of the relative time implies the absence of a
(near-)homogeneous magnetic field. In particular, this
means that odd-frequency superconductors exhibit a
diamagnetic Meissner effect even if no even-frequency
superconducting order is concurrently present.

There are additional fundamental open questions
in the context of odd-frequency pairing that require a
new perspective and where our approach could offer new
insights. For example, the relation of the anomalous
Green's functions  of Eq.~\eqref{eq:AnomCorr} that
describes the creation of a Cooper pair, and
\begin{align}\label{eq:AnomCorr_dagg}
\overline{F}_{\vek{i}\vek{j}}(t_1,t_2) = \langle T\,
c_\vek{i}(t_1)\, c_\vek{j}(t_2) \rangle
\end{align}
that stands for the annihilation of a pair, is
nontrivial~\cite{Heid1995,Solenov2009}. Using the Lehmann
or spectral representation of correlation functions,
adapted to anomalous correlation
functions~\cite{Gorkov1958,Abrikosov2012,Wang2012}, it
follows after Fourier transformation that $F_{\vek{i}
\vek{j}}\left(\omega\right)=\overline{F}^{*}_{\vek{j}
\vek{i}}\left(-\omega\right)$. On the other hand, for
states with well-defined symmetry under time exchange,
i.e., for either even- or odd-frequency superconductors,
it was shown in Ref.~\cite{Solenov2009} that $F_{\vek{i}
\vek{j}}\left(\omega\right)=\overline{F}^{*}_{\vek{j}
\vek{i}}\left(\omega\right)$ yields consistent solutions.
Only if the second relation holds is the thermodynamic
stability of  odd-frequency pairing at a continuous
transition guaranteed, and the superfluid stiffness
possesses the correct sign. While both expressions agree
for even-frequency pairing, they are inconsistent for
odd-frequency pairing. Even though a consistency argument
was made in Ref.~\cite{Solenov2009}, there exists no
proof that the second relation is correct for a given
microscopic Hamiltonian. If it is, one might have to
revise the spectral representation of the Gor'kov
function~\cite{Gorkov1958,Abrikosov2012,Wang2012}, even
for systems of infinite size. Addressing these crucial
issues ultimately requires a particle-number-conserving
description as presented in this manuscript.

Finally, having a particle-conserving
description of macroscopic pairing order is extremely
important if one wants to employ numerical approaches,
such as density matrix renormalization group or
Monte-Carlo simulations, which are performed for a fixed
particle number~\cite{Frick1992,Guerrero2000,Sorella2002,
Anderson2004,Astrakharchik2016,Ebling2021,Wietek2022}. With our
approach, we offer a first direct and unbiased way to
probe odd-frequency superconductivity using these
numerical methods.

To provide the reader with a general outline and
motivation for the main part of this article, the
fundamentals of our theoretical formalism and the
obtained results are briefly summarized in
Sec.~\ref{sec:outline}. This is followed by a thorough
development of the particle-conserving theory describing
time-dependent pair correlations, presented in
Sec.~\ref{sec:Formalism}. The rigorous formalism
discussed in Sec.~\ref{sec:Formalism} rests on minimal
assumptions about basic properties of the physical
system and is therefore widely applicable.
As a first demonstration of its utility, a fully general
derivation of the Meissner effect for odd-frequency
superconductors is presented in Sec.~\ref{sec:Meissner}.
Section~\ref{sec:Examples} is devoted to further
illustrating the formalism by its application to two
generic systems exhibiting symmetric-pairing order. The
spin-balanced \textit{s}-wave Fermi superfluid
subject to Zeeman splitting considered in
Sec.~\ref{sec:ExamplesTransf} serves as an example for
the case where symmetric pair correlations arise in
conjunction with conventional (antisymmetric-pairing,
even-frequency) superfluidity. We refer to this
situation as the \emph{transformer scenario}. In
contrast, the emergence of symmetric-pairing
correlations in the absence of ordinary superfluidity is
referred to as the \emph{generator scenario}. We
elucidate this alternative situation using the
composite-boson condensate as an example
(Sec.~\ref{sec:ExampleGen}). Our conclusions are
presented in Sec.~\ref{sec:Conc}. Details of some
mathematical derivations are provided in Appendices.

\section{Outline of theoretical approach and overview of
main results}\label{sec:outline}

A system of $N$ fermions realizes a Cooper-pair
condensate when the two-body reduced density matrix
$\rho_{\vek{i}\vek{j},\vek{k}\vek{l}} = \langle
c^\dagger_\vek{i} c^\dagger_\vek{j} c_\vek{l} c_\vek{k}
\rangle$ has an eigenvalue of order $N$~\cite{Yang1962},
which is generally referred to as a \emph{macroscopic}
eigenvalue~\cite{Leggett2006}. In this case, the density
matrix factorizes to leading order in $N$;
\begin{equation} \label{eq:RhoFact}
\rho_{\vek{i} \vek{j},\vek{k}\vek{l}} = \phi_{\vek{i}
\vek{j}}\, \phi^*_{\vek{k} \vek{l}} +
\tilde{\rho}_{\vek{i} \vek{j},\vek{k}\vek{l}} \quad ,
\end{equation}
where the dominant eigenvector $\phi_{\vek{i}\vek{j}}$
is the pair-condensate order parameter with eigenvalue
$n_0 = \sum_{\vek{i}\vek{j}} |\phi_{\vek{i}\vek{j}}|^2
\sim \mathcal{O}(N)$. The remaining part
$\tilde{\rho}_{\vek{i} \vek{j},\vek{k}\vek{l}}$ of the
two-body reduced density matrix has no macroscopic
eigenvalue. The pair-condensate order parameter
$\phi_{\vek{i}\vek{j}}$ is the fixed-$N$ analog of the
anomalous pair correlator $F_{\vek{i} \vek{j}}(0,0)$ of
Eq.~(\ref{eq:AnomCorr}) at equal time. Both
$\phi_{\vek{i}\vek{j}}$ and $F_{\vek{i} \vek{j}}$ may
equally serve as a starting point for the approximate
description of superfluidity and
superconductivity~\cite{Leggett2006}, but only 
$\phi_{\vek{i}\vek{j}}$ is well defined (and finite)
when the number of particles is fixed.

In this work, we study the properties of the
time-dependent two-body correlation matrix (T2bCM)
\begin{align}\label{eq:T2bCMgen}
\rho_{\vek{i}\vek{j},\vek{k}\vek{l}}(t_1,t_2) = \Langle
c^\dagger_{\vek{i}}(t_1)\, c^\dagger_{\vek{j}}(t_2)\,
c_{\vek{l}}(t_2)\, c_{\vek{k}}(t_1) \Rangle \quad .
\end{align}
This two-particle correlation function with a specific
choice of the time arguments serves as a useful
generalisation of the two-body reduced density matrix,
to which it reduces when $t_1=t_2=0$. Because the T2bCM
is a Hermitian and positive-semidefinite matrix in the
fermion-pair index space (see
Sec.~\ref{subsec:2bRDMprops}), we know that its
eigenvalues are real and non-negative, and its
eigenvectors are orthogonal. The T2bCM is a natural
quantity for introducing time dependence into the
description of superfluidity and superconductivity in a
number-conserving formalism. We show that the T2bCM
provides a general framework for the study of
odd-frequency pairing of fermions.

Assuming systems governed by a time-independent and
Hermitian Hamiltonian, the T2bCM depends only on the
relative time $t=t_1 - t_2$. A macroscopic pairing order
is signified by the presence of a macroscopic eigenvalue
of the  T2bCM. In this case, we find  a time-dependent
factorization to leading order in $N$,
\begin{equation}\label{eq:T2bCMfact}
\rho_{\vek{i} \vek{j},\vek{k}\vek{l}}(t,0) =
\phi_{\vek{i}\vek{j}}(t)\, \phi^*_{\vek{k} \vek{l}}(t) +
\tilde{\rho}_{\vek{i} \vek{j},\vek{k}\vek{l}}(t)\quad ,
\end{equation}
which is analogous to the time-independent case of
Eq.~(\ref{eq:RhoFact}). The time-dependent dominant
eigenvalue is $n_0(t) =\sum_{\vek{i}\vek{j}}
|\phi_{\vek{i}\vek{j}}(t)|^2 \sim \mathcal{O}(N)$, and
the remaining part $\tilde{\rho}_{\vek{i}\vek{j},
\vek{k}\vek{l}}(t, 0)$ has nonmacroscopic eigenvalues
$\sim\mathcal{O}(N^0)$ or smaller. The time-dependent
pair-condensate order parameter $\phi_{\vek{i}\vek{j}}
(t)$ is the fixed-$N$ analog of the anomalous pair
correlator $F_{\vek{i}\vek{j}}(t,0)$ at relative time
$t$. Equation~\eqref{eq:T2bCMfact} is the starting point
for our study of odd-frequency pairing of fermions. It
is reminiscent of the factorization of the two-body
correlation function postulated by
Gorkov~\cite{Gorkov1958} in the context of
superconductivity, but it is rigorous and applies to any
finite and number-conserving system. 

In order to study odd-frequency superfluidity, we
introduce the symmetric and antisymmetric parts of the
time-dependent pair-condensate order parameter 
\begin{equation}
\phi_{\vek{i}\vek{j}}(t) = \phi^\mathrm{(a)}_{\vek{i}
\vek{j}}(t) + \phi^\mathrm{(s)}_{\vek{i}\vek{j}}(t)
\quad ,
\end{equation}
where $\phi^\mathrm{(a)}_{\vek{i}\vek{j}}(t) =
-\phi^\mathrm{(a)}_{\vek{j}\vek{i}}(t)$ is antisymmetric
under index exchange $\vek{i}\leftrightarrow \vek{j}$,
and $\phi^\mathrm{(s)}_{\vek{i}\vek{j}}(t) =
\phi^\mathrm{(s)}_{\vek{j}\vek{i}}(t)$ is symmetric. For
the time-ordered correlation function $F_{\vek{i}\vek{j}}
(t_1,t_2)$ of Eq.~(\ref{eq:AnomCorr}), even symmetry
under the exchange of fermion indices is equivalent to
odd symmetry in the relative time due to the properties
of the time-ordering operator. We thus take the
emergence of a macroscopic symmetric part
$\phi^\mathrm{(s)}_{\vek{i}\vek{j}}(t)\sim \mathcal{O}
(\sqrt{N})$ as synonymous with the existence of
odd-frequency pairing order. If a macroscopic
antisymmetric part $\phi^\mathrm{(a)}_{\vek{i}\vek{j}}
(t)$ exists, it indicates the presence of (conventional)
even-frequency pairing order. 

The time dependence of the pair-condensate order
parameter is also important. Specifically, at relative
time $t=0$, it  is purely antisymmetric, $\phi_{\vek{i}
\vek{j}}(0) = -\phi_{\vek{j}\vek{i}}(0)$. A nonzero
value of the symmetric part $\phi^\mathrm{(s)}_{\vek{i}
\vek{j}}(t)$ can only develop at $t\ne 0$. Thus, we can
distinguish two distinct scenarios for the presence of
macroscopic symmetric-pairing (i.e., odd-frequency)
order, depending on whether a nonzero and macroscopic
$\phi^\mathrm{(a)}_{\vek{i}\vek{j}}(0)$ exists.
 
\textit{Coexistence of symmetric and antisymmetric
pairing---Transformer scenario}.
A finite $\phi^\mathrm{(a)}_{\vek{i}\vek{j}}(0)\equiv
\phi_{\vek{i}\vek{j}}$ signals the presence of an
ordinary Cooper-pair condensate. At finite relative time
$t$, a macroscopic symmetric-pairing order can coexist
with the ordinary Cooper-pair condensate. A small-$t$
expansion reveals that the symmetric part of the
pair-condensate order parameter is related to the
conventional Cooper-pair order parameter $\phi_{\vek{i}
\vek{j}}$ by a transformation matrix,
\begin{subequations}
\begin{align} \label{eq:TransMatFirstOrder}
\phi^\mathrm{(s)}_{\vek{i}\vek{j}}(t) &= -i\,
\frac{t}{\hbar}\, \frac{1}{n_0} \sum_{\vek{k}\vek{l}}
\tau_{\vek{i}\vek{j},\vek{k}\vek{l}}\, \phi_{\vek{k}
\vek{l}} + \mathcal{O}(t^2) \quad , \\
\tau_{\vek{i} \vek{j},\vek{k}\vek{l}} &= \frac{1}{2}\, 
\Langle \big(c^\dagger_\vek{i}\, H\,c^\dagger_\vek{j} +
c^\dagger_\vek{j}\, H\,c^\dagger_\vek{i} \big)\,
c_\vek{l} c_\vek{k} \Rangle  \quad . \label{eq:TransMat}
\end{align}          
\end{subequations}
This transformation matrix evidently depends not only on
the properties of the underlying quantum state, but also
on the Hamiltonian $H$ of the system. We refer to this
situation as the \emph{transformer scenario}. We derive
more-general expressions for arbitrary time dependence
in Sec.~\ref{subsubsec:TransMech} and illustrate the
transformer scenario using the spin-balanced
\textit{s}-wave Fermi superfluid with finite
Zeeman splitting as an example in
Sec.~\ref{sec:ExamplesTransf}.

\textit{Emergence of symmetric pairing from hidden
orders---Generator scenario}.
In the absence of a conventional Cooper-pair condensate,
i.e., when $\phi_{\vek{i}\vek{j}}(0)\equiv\phi_{\vek{i}
\vek{j}}=0$, macroscopic pairing order can only emerge
at $t\ne 0$ and is embodied in an order parameter whose
leading time dependence is linear in $t$,
\begin{align}\label{eq:GenOPintro}
\phi_{\vek{i}\vek{j}}(t) = \frac{t}{\sqrt{2}}\,
\sqrt{n_0^{\prime\prime}(0)}\, \chi_{0,\vek{i}\vek{j}} +
\mathcal{O}(t^2) \quad ,
\end{align}
where $n_0^{\prime\prime}(0)$ is the second derivative
of the dominant eigenvalue with respect to $t$, and
$\chi_{0,\vek{i}\vek{j}}$ are the elements of a
normalized eigenvector of the T2bCM at $t=0$.
Symmetric-pairing order is present when the order
parameter (\ref{eq:GenOPintro}) has a nonvanishing
symmetric part $\phi_{\vek{i}\vek{j}}^{(\mathrm{s})}(t)
\ne 0$. We refer to this situation as the
\emph{generator scenario}. In the special case when the
order parameter is fully symmetric, i.e., $\phi_{\vek{i}
\vek{j}}(t)=\phi_{\vek{i}\vek{j}}^{(\mathrm{s})}(t)$,
the eigenvalue equation determining $n_0^{\prime\prime}
(0)$ and $\chi_{0,\vek{i}\vek{j}}$ simplifies to
$\sum_{\vek{k}\vek{l}}\gamma_{\vek{i}\vek{j},\vek{k}
\vek{l}}\, \chi_{0,\vek{k}\vek{l}} = [\hbar^2 \,
n_0^{\prime\prime}(0)/2]\,\chi_{0,\vek{i}\vek{j}}$, with
generator matrix elements
\begin{align}
\gamma_{\vek{i}\vek{j},\vek{k}\vek{l}} &= \frac{1}{4}
\Langle \big(c^\dagger_\vek{i}\, H\, c^\dagger_\vek{j} +
c^\dagger_\vek{j}\, H\, c^\dagger_\vek{i} \big)\, \big(
c_\vek{l}\, H\, c_\vek{k}  + c_\vek{k}\, H\, c_\vek{l}
\big)\Rangle \nonumber\\
& \hspace{5.3cm} +\,\mathcal{O}\big(N^0\big) \,\, .
\end{align}
These results are derived in a more general setting in
Sec.~\ref{subsubsec:GenMech}. Normally, a macroscopic
$n_0^{\prime\prime}(0)$ can arise if the underlying 
quantum state has a composite, multi-particle
condensate. Such a hidden order is not visible in the
conventional two-body reduced density matrix. We
illustrate the generator scenario in
Sec.~\ref{sec:ExampleGen}, using the composite-boson
condensate formed by Cooper pairs coupled to magnons as
an example.

The transformer and generator scenarios exhaust all
possibilities for symmetric-pairing order with leading
linear-in-$t$ dependence to emerge in any system. Thus,
analyzing the structure of the transformer and generator
matrices for a particular physical situation provides
the means to identify necessary and sufficient
conditions under which odd-frequency superfluidity may
be realized. Our theory could be extended to discuss the 
potential for unconventional pairing orders to manifest
via a higher-order $t$ dependence.

Using the generalized Penrose-Onsager-Yang formalism, we
show that macroscopic quantum coherence in the T2bCM
implies the conventional (diamagnetic) Meissner effect
in both the transformer and generator scenarios.

\section{Penrose-Onsager-type formalism for
time-dependent pair correlations}\label{sec:Formalism}

\subsection{Basic definitions and assumptions}
\label{subsec:Defs&Assumps}

Our starting point is the time-dependent two-particle
correlation function (four-point function) with a
specific choice for the time arguments. 
Equation~(\ref{eq:T2bCMgen}) defines the matrix elements
of the time-dependent two-body correlation matrix
(T2bCM) $\bbarb{\rho}(t_1,t_2)$. In this article, we
generally denote a matrix in two-particle index space
that has matrix elements $M_{\vek{i}\vek{j},\vek{k}
\vek{l}}$ by the symbol $\bbarb{M}$. Each individual
index $\vek{i}$ labels a single-particle state that is
created [annihilated] at time $t$ by its corresponding
fermion operator $c^\dagger_{\vek{i}}(t)$ [$c_{\vek{i}}
(t)$]. An index pair $\vek{i}\vek{j}$ refers to the
fermionic two-particle state where one of the fermions
is in state $\vek{i}$ and the other in state $\vek{j}$. 

The expectation value in Eq.~(\ref{eq:T2bCMgen}) is to
be taken with respect to the system's many-body ground
state (if we consider the zero-temperature limit) or a
thermal mixture of many-body states (when considering a
system at finite temperature). Since we are interested
in describing superfluidity from a particle-conserving
perspective, we furthermore restrict ourselves, in the
zero-temperature case, to situations where the ground
state is an eigenstate of the fermion-number operator
$\hat{N} =\sum_\vek{i} c^\dagger_\vek{i}c_\vek{i}$ with
eigenvalue $N$. [Here and in the following, we use the
shorthand notation where $c_\vek{i}\equiv c_\vek{i}(0)$.]
In the finite-temperature case, we work over statistical
ensembles in which the microstates all have the same
particle number $N$. We also assume that the Hamiltonian
$H$ itself commutes with $\hat{N}$. These assumptions
underlie the majority of the discussion in this article,
and it will be explicitly stated whenever we deviate
from them.

Vectors in two-particle index space are denoted as
$\barb{f}$ and have components $f_{\vek{i}\vek{j}}$.
Vectors that are invariant (invariant up to a minus
sign) under the particle-index exchange $\vek{i}
\leftrightarrow \vek{j}$ are called symmetric
(antisymmetric). An arbitrary vector can be projected
onto its symmetric and antisymmetric parts by applying
the projectors $\bbarb{S}$ and $\bbarb{A}$ respectively,
which have the matrix elements
\begin{subequations}\label{eq:SAdefs}
\begin{align}
S_{\vek{i}\vek{j},\vek{k}\vek{l}} &= \frac{1}{2}\big(
\delta_{\vek{i},\vek{k}}\,\delta_{\vek{j},\vek{l}} +
\delta_{\vek{i},\vek{l}}\,\delta_{\vek{j},\vek{k}}\big)
\quad , \label{eq:Sdef}\\[0.1cm]  
A_{\vek{i}\vek{j},\vek{k}\vek{l}} &= \frac{1}{2}\big(
\delta_{\vek{i},\vek{k}}\,\delta_{\vek{j},\vek{l}} -
\delta_{\vek{i},\vek{l}}\,\delta_{\vek{j},\vek{k}}\big)
\quad . \label{eq:Adef}
\end{align}
\end{subequations}
Here $\delta_{\vek{a},\vek{b}}$ is a multidimensional
Kronecker delta function with discrete vector arguments
$\vek{a}$ and $\vek{b}$. The definitions in
Eqs.~(\ref{eq:SAdefs}) yield the following properties of
the projector matrices:
\begin{align}\label{eq:SAprops}
\bbarb{A} + \bbarb{S} = \bbarb{1} \,\,\, ; \,\,\,
\bbarb{A}^2 = \bbarb{A} \,\,\, ; \,\,\, \bbarb{S}^2 =
\bbarb{S} \,\,\, ; \,\,\, \bbarb{A}\bbarb{S} =
\bbarb{S}\bbarb{A} = \bbarb{0} \,\,\, ,
\end{align}
where $\bbarb{0}$ and $\bbarb{1}$ are the zero and unit
matrices in two-particle index space, respectively.

\subsection{The two-body reduced density matrix}
\label{subsec:2bRDM}

Setting all time arguments in Eq.~(\ref{eq:T2bCMgen}) to
zero shows that $\bbarb{\rho}(0,0) \equiv \bbarb{\rho}$,
with its matrix elements
\begin{align}\label{eq:2bRDMdef}
\rho_{\vek{i}\vek{j},\vek{k}\vek{l}}(0,0) \equiv
\rho_{\vek{i} \vek{j},\vek{k}\vek{l}} = \Langle
c^\dagger_\vek{i} c^\dagger_\vek{j} c_\vek{l} c_\vek{k}
\Rangle \,\, ,
\end{align}
corresponds to the two-body reduced density matrix
(2bRDM), which is central to the standard discussion of
Cooper-pair condensation~\cite{Yang1962,Leggett2006}.

The 2bRDM is Hermitian and positive-semidefinite. For
systems where the total particle number $N$ is fixed,
the trace evaluates to $\Tr\,\bbarb{\rho} = N(N-1)$.
As a consequence, the eigenvalues of the 2bRDM are
non-negative and sum to the constant $N(N-1)$; hence,
they can be interpreted as occupation numbers of
fermion-pair states that relate to the eigenvectors of
$\bbarb{\rho}$. Yang~\cite{Yang1962} further showed that
the eigenvalues are bounded from above by $N$. Following
Leggett~\cite{Leggett2006}, and based on the ideas of
Penrose, Onsager and Yang~\cite{Penrose1956,Yang1962},
we define pseudo-Bose-Einstein condensation (pseudo-BEC)
of fermion (Cooper) pairs by the presence of at least
one eigenvalue that is macroscopic, i.e., of order $N$.
This becomes a rigorous definition in the thermodynamic
limit, i.e., when the number of particles $N$ is taken
to infinity while keeping the fermion number density
constant. 

We will usually be concerned with the situation where
only a single eigenvalue of $\bbarb{\rho}$ is
macroscopic, realizing a simple pseudo-BEC~\footnote{In
principle, there could be multiple macroscopic
eigenvalues of the 2bRDM, corresponding to a fragmented
pseudo-BEC, but this situation appears to be rare in
nature~\cite{Leggett2006}.}. The fact that the sum over
all eigenvalues is $\mathcal{O}(N^2)$ implies that there
are typically many remaining eigenvalues of order unity.
The pseudo-BEC has to be contrasted with the situation
where all the eigenvalues are of order unity, which
would represent an uncondensed system configuration with
no macroscopic order~\cite{Penrose1956,Yang1962}.

\subsection{General properties of the time-dependent
two-body correlation matrix}\label{subsec:2bRDMprops}

Evidently, the hermiticity, positive-semidefiniteness
and trace properties of the 2bRDM $\bbarb{\rho}\equiv
\bbarb{\rho}(0,0)$ are instrumental in drawing
conclusions about the macroscopicity of its eigenvalues
and, thus, the possibility of time-independent
conventional macroscopic order. It is now demonstrated
that the T2bCM $\bbarb{\rho}(t_1, t_2)$ defined in
Eq.~(\ref{eq:T2bCMgen}) satisfies these same three
properties, even for nonzero and generally distinct time
arguments $t_1$ and $t_2$.

\textit{(i)~Hermiticity}: We have 
\begin{align}\label{eq:ProofHerm}
\rho^*_{\vek{k}\vek{l},\vek{i}\vek{j}}(t_1,t_2) &=
\Langle c^\dagger_\vek{i}(t_1)\,c^\dagger_\vek{j}(t_2)\,
c_\vek{l}(t_2)\,c_\vek{k}(t_1) \Rangle \nonumber \\
&= \rho_{\vek{i}\vek{j},\vek{k}\vek{l}}(t_1, t_2) \,\, ,
\end{align}
so that $\bbarb{\rho}^\dagger(t_1,t_2) = \bbarb{\rho}
(t_1, t_2)$ for all $t_1$ and $t_2$. Thus, the T2bCM
(\ref{eq:T2bCMgen}) is Hermitian.

\textit{(ii)~Positive-semidefiniteness}: For an
arbitrary vector $\barb{f}$ in two-particle index space,
we have
\begin{align}\label{eq:ProofSemiDef}
\barb{f}^\dagger\,\bbarb{\rho}(t_1,t_2)\,\barb{f} &=
\sum_{\vek{i}\vek{j},\vek{k}\vek{l},\alpha} \,f_{\vek{i}
\vek{j}}^* \, \langle c^\dagger_\vek{i}(t_1)\,
c^\dagger_\vek{j}(t_2)\ket{\alpha} \nonumber \\
& \hspace{2.5cm} \times \bra{\alpha} c_\vek{l}(t_2)\,
c_\vek{k}(t_1)\rangle\, f_{\vek{k}\vek{l}} \nonumber
\\[0.2cm]
&= \sum_\alpha \left| \sum_{\vek{i}\vek{j}} f^*_{\vek{i}
\vek{j}}\,\langle c^\dagger_\vek{i}(t_1)\,
c^\dagger_\vek{j}(t_2)\ket{\alpha}\right|^2 \geq 0 \,\, ,
\end{align}
where a resolution of the identity in the
$(N-2)$-particle Hilbert space was inserted. Thus,
$\bbarb{\rho}(t_1, t_2)$ is positive-semidefinite, and
its eigenvalues are always non-negative.

\textit{(iii)~Time-independent trace}: Lastly, for a
system with fixed particle number $N$, and a
number-conserving Hamiltonian $H$, we have $\hat{N}(t) =
\sum_\vek{i} c^\dagger_\vek{i}(t)\,c_\vek{i}(t) =
\hat{N}(0) \equiv \hat N$. It follows that the trace of
the T2bCM is equal to the time-independent constant
$N(N-1)$: 
\begin{align}\label{eq:ProofTrace}
\Tr\bbarb{\rho}(t_1,t_2) &= \sum_{\vek{i}\vek{j}} \Langle
c^\dag_\vek{i}(t_1)\,c^\dag_\vek{j}(t_2)\,c_\vek{j}(t_2)
\, c_\vek{i}(t_1) \Rangle \nonumber \\
&= \sum_\vek{i} \Langle c^\dag_\vek{i}(t_1)\, \hat{N}\,
c_\vek{i}(t_1) \Rangle =  N(N-1) \,\, .
\end{align}
    
Given points (i), (ii) and (iii) above, it is assured
that the T2bCM $\bbarb{\rho}(t_1,t_2)$ has non-negative
eigenvalues that sum to $N(N-1)$. Thus, the
interpretation of the eigenvalues as occupation numbers
of generalized fermion-pair orbitals is also valid for
the T2bCM.

In the following, we assume that the system's time
evolution is determined by a Hermitian time-independent
Hamiltonian $H$ that commutes with the fermion-number
operator $\hat{N}$, and that the system is in an
equilibrium state with fixed particle number $N$. This
may be the ground state of $H$ or a thermal mixture. In
this case, the T2bCM depends only on the relative time
$t = t_1 - t_2$, and we will denote it by $\bbarb{\rho}
(t) \equiv \bbarb{\rho}(t,0)$. The matrix elements of
this quantity are then given by
\begin{align}\label{eq:T2bCMdef}
&\rho_{\vek{i}\vek{j},\vek{k}\vek{l}}(t) = \Langle
c^\dagger_\vek{i}\,\exp\big( \tfrac{-i\, t}{\hbar}\, H
\big)\,c^\dagger_\vek{j}\,c_\vek{l}\,\exp\big( \tfrac{i
\, t}{\hbar}\, H\big)\,c_\vek{k} \Rangle \,\, ,
\end{align}
specializing to the 2bRDM (\ref{eq:2bRDMdef}) when $t=0$;
$\bbarb{\rho}(0) \equiv \bbarb{\rho}$.
We note that expressions that we  derive for the
short-time expansion of this correlation function might
prove very useful for numerical approaches that want to
probe whether a given many-body state displays
odd-frequency pairing correlations.

\subsection{Index-exchange symmetry of the
time-dependent two-body correlation matrix}
\label{sec:tVsIndexX}

The number-conserving formalism leads us to reconsider
the question of what odd-frequency pairing correlations
mean. Let us first establish that permuting the
time-arguments in the pair correlation function of
Eq.~\eqref{eq:AnomCorr} is equivalent to permuting the
fermion operators with a change of sign, i.e.,
\begin{align}\label{eq:GorkovSymmetry}
F_{\vek{i}\vek{j}}(t_1,t_2) = -  F_{\vek{j}\vek{i}}
(t_2,t_1) \quad .
\end{align}
This follows directly from the definition of the
time-ordering operator $T$~\cite{Berezinskii1974,
Linder2019,Dahal2009}. As a consequence, any odd-in-time
component of the pair correlation function (equivalent to
odd-frequency in the frequency domain) is even under
exchange of the fermion indices. The search for
odd-in-time (or odd-frequency) pairing correlations is
thus equivalent to the search for
even-under-fermion-exchange pairing correlation. We
argue that the latter is a more natural way to think
about this type of unconventional pairing than the
symmetry of the pairing correlations in time or
frequency, and we will use this formulation in the
following. Moreover, the exchange-symmetry concept can be
meaningfully applied also to correlation functions that
are defined without time ordering, such as the T2bCM
introduced in Eq.~(\ref{eq:T2bCMgen}). As shown in our
analysis presented below, the T2bCM exhibits its own
particular correspondence between index-exchange
symmetry and time dependence. To be specific, we focus
our discussion on the $t$ dependence of $\bbarb{\rho}
(t)$ defined in Eq.~(\ref{eq:T2bCMdef}), reminding the
reader that $t$ denotes the relative time $t\equiv t_1
- t_2$.

\begin{figure}[t]
\includegraphics[width=0.95\columnwidth]{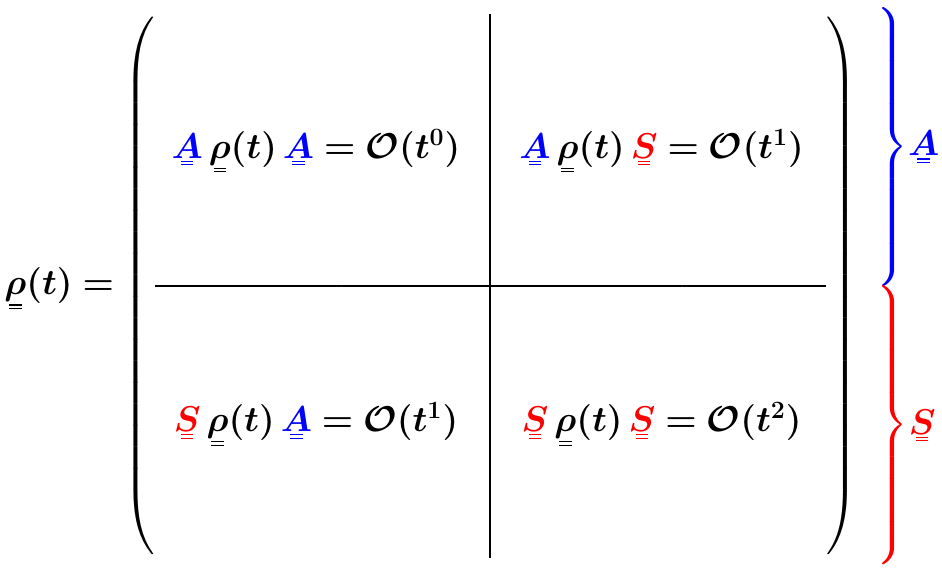}
\caption{\label{fig:blockRho} General structure of the
time-dependent two-body correlation matrix (T2bCM)
$\bbarb{\rho}(t)$. Here $\bbarb{A}$ and $\bbarb{S}$
denote the projectors onto subspaces spanned by
two-particle basis states that are antisymmetric and
symmetric, respectively, under particle exchange. See
Eqs.~(\ref{eq:SAdefs}).}
\end{figure}

The resolution of the identity in Eq.~(\ref{eq:SAprops})
can be used to decompose $\bbarb{\rho}(t)$ into its
parts that are either fully symmetric, fully
antisymmetric, or have mixed symmetry under index
exchange,
\begin{align}\label{eq:RhoDecomp}
\bbarb{\rho}(t) = \bbarb{A}\,\bbarb{\rho}(t)\,\bbarb{A}
+ \bbarb{A}\,\bbarb{\rho}(t)\,\bbarb{S} + \bbarb{S}\,
\bbarb{\rho}(t)\,\bbarb{A} + \bbarb{S}\,\bbarb{\rho}(t)
\,\bbarb{S} \,\,\, .
\end{align}
It is easy to show (see the previous section) that
both matrices $\bbarb{A}\,\bbarb{\rho}(t)\,\bbarb{A}$
and $\bbarb{S}\,\bbarb{\rho}(t)\,\bbarb{S}$ are
Hermitian and positive-semidefinite, thus have real and
non-negative eigenvalues. Furthermore, $\Tr \bbarb{A}\,
\bbarb{\rho}(t)\,\bbarb{A} + \Tr \bbarb{S}\,\bbarb{\rho}
(t)\,\bbarb{S} = \Tr\bbarb{\rho}(t)=N(N-1)$. At $t = 0$,
all terms in Eq.~(\ref{eq:RhoDecomp}) containing the
projector $\bbarb{S}$ vanish because $\bbarb{S}\,
\bbarb{\rho}(0)=\bbarb{\rho}(0)\,\bbarb{S} = \bbarb{0}$
due to the anticommutation relations satisfied by
fermion operators. Thus, the only nonzero contribution
to the 2bRDM $\bbarb{\rho}(0)$ is the purely
antisymmetric block;
\begin{align}\label{eq:Rho(0)}
\bbarb{\rho}(0) = \bbarb{A}\,\bbarb{\rho}(0)\,\bbarb{A}
\quad .
\end{align}
The part in Eq.~(\ref{eq:RhoDecomp}) that is fully
symmetric under index exchange, as well as the parts
that have mixed symmetry, only emerge at $t\ne 0$. To
illustrate this more clearly, it is useful to consider
the small-$t$ expansion
\begin{align}\label{eq:RhoTimeExp}
\bbarb{\rho}(t) = \bbarb{\rho}(0) + t\,
\bbarb{\rho}^{\prime}(0) + \frac{t^2}{2}\,
\bbarb{\rho}^{\prime\prime}(0)+\mathcal{O}(t^3) \quad ,
\end{align}
where the terms $\bbarb{\rho}^{\prime}(0)$ and
$\bbarb{\rho}^{\prime\prime}(0)$ have matrix elements
given by
\begin{subequations}\label{eq:Prime&PrimePrimeDefs}
\begin{align}
\rho^{\prime}_{\vek{i}\vek{j},\vek{k}\vek{l}}(0) &=
\partial_t\,\rho_{\vek{i}\vek{j},\vek{k}\vek{l}}(t)\,
\big|_{t = 0} \nonumber\\
&= \frac{i}{\hbar}\left( \Langle c^\dagger_\vek{i}
c^\dagger_\vek{j} c_\vek{l} H c_\vek{k} \Rangle -
\Langle c^\dagger_\vek{i} H c^\dagger_\vek{j} c_\vek{l}
c_\vek{k} \Rangle \right) \,\,\, , \label{eq:RhoPrime}
\\[0.30cm]
\rho^{\prime\prime}_{\vek{i}\vek{j},\vek{k}\vek{l}}(0)
&= \partial^2_t\,\rho_{\vek{i}\vek{j},\vek{k}\vek{l}}(t)
\,\big|_{t = 0} \nonumber\\
&= \frac{1}{\hbar^2} \left( 2\,\Langle
c^\dagger_\vek{i} H c^\dagger_\vek{j} c_\vek{l} H
c_\vek{k} \Rangle - \Langle c^\dagger_\vek{i} H^2
c^\dagger_\vek{j} c_\vek{l} c_\vek{k} \Rangle \right.
\nonumber\\
& \left. \hspace{3cm} -\,\,\Langle c^\dagger_\vek{i}
c^\dagger_\vek{j} c_\vek{l} H^2 c_\vek{k} \Rangle
\right) \,\,\, . \label{eq:RhoPrimePrime}
\end{align}
\end{subequations}
Inspection of Eqs.~(\ref{eq:Prime&PrimePrimeDefs})
reveals that $\bbarb{\rho}^{\prime}(0)$ and
$\bbarb{\rho}^{\prime\prime}(0)$ obey the symmetry
constraints
\begin{subequations}\label{eq:Prime&PrimePrimeSymm}
\begin{align}
\bbarb{\rho}^{\prime}(0) &=
\bbarb{A}\,\bbarb{\rho}^{\prime}(0)\,\bbarb{A} +
\bbarb{A}\,\bbarb{\rho}^{\prime}(0)\,\bbarb{S} +
\bbarb{S}\,\bbarb{\rho}^{\prime}(0)\,\bbarb{A} \,\,\, ,
\label{eq:RhoPrimeSymm} \\[0.20cm]
\bbarb{\rho}^{\prime\prime}(0) &=
\bbarb{A}\,\bbarb{\rho}^{\prime\prime}(0)\,\bbarb{A} +
\bbarb{A}\,\bbarb{\rho}^{\prime\prime}(0)\,\bbarb{S} +
\bbarb{S}\,\bbarb{\rho}^{\prime\prime}(0)\,\bbarb{A} +
\bbarb{S}\,\bbarb{\rho}^{\prime\prime}(0)\,\bbarb{S}
\,\,\, . \label{eq:RhoPrimePrimeSymm}
\end{align}    
\end{subequations}
Therefore, the small-$t$ limit for the fully symmetric
block of the T2bCM is $\bbarb{S}\,\bbarb{\rho}(t)\,
\bbarb{S} = t^2 \bbarb{S}\,\bbarb{\rho}^{\prime\prime}(0)
\,\bbarb{S}\, /2 + \mathcal{O}(t^3)$, while the
mixed-symmetry blocks $\bbarb{A}\,\bbarb{\rho}(t)\,
\bbarb{S}$ and $\bbarb{S}\,\bbarb{\rho}(t)\,\bbarb{A}$
are $\mathcal{O}(t)$. See Fig.~\ref{fig:blockRho} for an
illustration of this general structure. The relationship
between index-exchange symmetry and leading-order $t$
dependence signals the relevance of certain blocks of
the T2bCM for the description of odd-frequency pairing
correlations.

\subsection{Index-exchange symmetry of eigenvectors}
\label{sec:tVsIndexEig}

Our discussion in the previous Sec.~\ref{sec:tVsIndexX}
elucidated the connection between index-exchange symmetry
and $t$ dependence of the T2bCM $\bbarb{\rho}(t)$. In
particular, it was shown that $\bbarb{\rho}(0)$ is
purely antisymmetric, as required for a 2bRDM, and that
the part $\bbarb{S}\,\bbarb{\rho}(t)\,\bbarb{S}$ is
$\mathcal{O}(t^2)$ in the limit $t\to 0$. We now explore
ramifications of this structure for the eigenvectors of
the T2bCM, which are essential for describing pairing
correlations present in the system~\cite{Leggett2006}.
Insights obtained here about general properties of T2bCM
eigenvectors underpin the description of macroscopic
symmetric-pairing order in terms of a single eigenvector
with macroscopic eigenvalue, developed in the subsequent
Sec.~\ref{sec:SymmOP}.

We start with the spectral decomposition
\begin{align}\label{eq:Rho(t)SpecDecomp}
\bbarb{\rho}(t) = \sum_\alpha n_\alpha(t)\,
\barb{\chi}_\alpha(t)\, \barb{\chi}^\dagger_\alpha(t)
\quad , 
\end{align}
where the $n_\alpha(t)$ are eigenvalues of the T2bCM,
and $\barb{\chi}_\alpha(t)$ the corresponding normalised
eigenvectors. The eigenvalues $n_\alpha(0)$ of the 2bRDM
$\bbarb{\rho}(0)$ are usually interpreted as occupation
numbers for two-fermion states described by
wavefunctions $\barb{\chi}_\alpha
(0)$~\cite{Leggett2006}. In fact, as shown in more
detail in Appendix~\ref{app:recurse}, the vectors
$\barb{\chi}_\alpha(0)$ with nonzero eigenvalue
$n_\alpha(0)$ are fully antisymmetric;
$\barb{\chi}_\alpha(0)\equiv\bbarb{A}\,\barb{\chi}_\alpha
(0)$, and so possess the crucial property of two-fermion
wavefunctions to be antisymmetric under index exchange
$\vek{i}\leftrightarrow \vek{j}$, required by Fermi
statistics. As was demonstrated in
Sec.~\ref{subsec:2bRDMprops}, the eigenvalues
$n_\alpha(t)$ of the T2bCM defined in
Eq.~(\ref{eq:T2bCMdef}) are also non-negative and can
therefore be interpreted as occupation numbers for the
generalized natural pair orbitals $\barb{\chi}_\alpha
(t)$. However, due to the form of
Eq.~(\ref{eq:RhoDecomp}), the eigenvectors
$\barb{\chi}_\alpha(t)$ for $t\ne 0$ are not constrained
to have identically zero symmetric part and, therefore,
can encode symmetric pair correlations that do not
correspond to any ordinary two-fermion bound state. One
of the important observations resulting from the
number-conserving formalism is that there are two
distinct mechanisms for symmetric components $\bbarb{S}\,
\barb{\chi}_\alpha(t)$ of T2bCM eigenvectors
$\barb{\chi}_\alpha(t)$ to exist. We elucidate these two
mechanisms in turn below. 

\begin{figure*}
\includegraphics[width=0.80\textwidth]{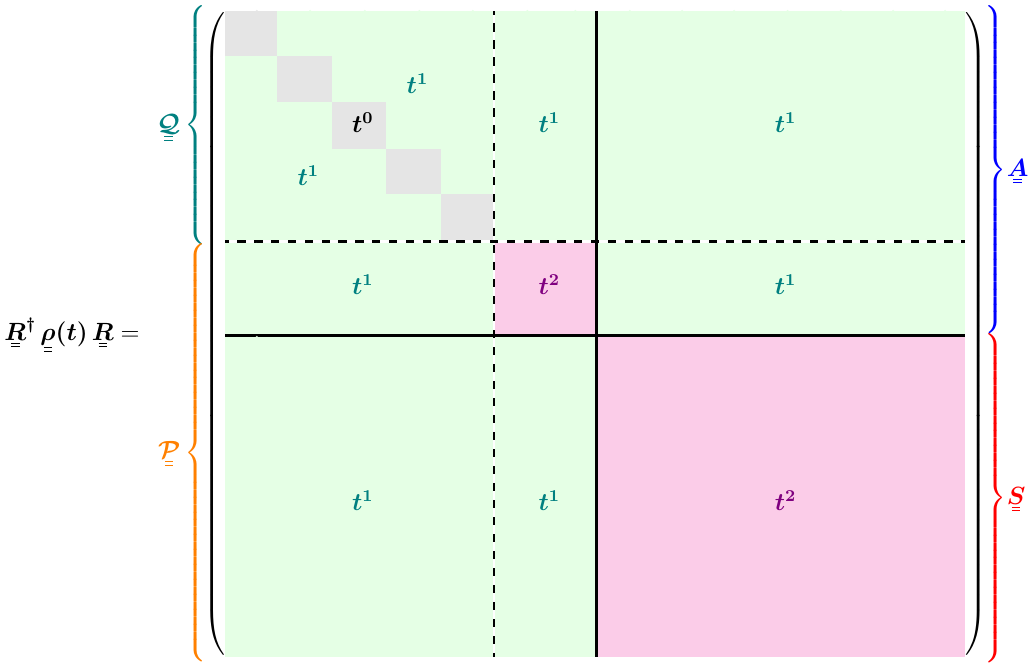}
\caption{\label{fig:rhoTrans} Structure of the
time-dependent two-body correlation matrix (T2bCM)
$\bbarb{\rho}(t)$ in the representation where
$\bbarb{\rho}(0)$ is diagonal. The form shown here is
related to the form shown in Fig.~\ref{fig:blockRho} by
a unitary transformation $\bbarb{R}$, where the columns
of $\bbarb{R}$ are the normalised eigenvectors of
$\bbarb{\rho}(0)$. Labels $t^n$ in the various blocks
indicate their leading-order small-$t$ dependence. The
operators $\bbarb{A}$ and $\bbarb{S}$ project onto
subspaces of two-particle states that are antisymmetric
and symmetric, respectively, under particle exchange.
The subspace where $\bbarb{\rho}(0)$ has positive
eigenvalues is singled out by the projector $\Cproj$,
and $\Dproj=\bbarb{1}-\Cproj$ projects onto the
nullspace of $\bbarb{\rho}(0)$.}
\end{figure*}

\subsubsection{Transformer mechanism}
\label{subsubsec:TransMech}

The transformer mechanism applies to T2bCM eigenvectors
$\barb{\chi}_\alpha(t)$ with a nonzero eigenvalue
$n_\alpha(0)>0$ at $t=0$. In this case, the eigenvector
belongs to the $\bbarb{A}\,\bbarb{\rho}\,\bbarb{A}$ block
and satisfies the eigenvalue equation of the 2bRDM
\begin{equation}\label{eq:2bRDMeigen}
\bbarb{\rho}\, \barb{\chi}_\alpha(0) = n_\alpha(0)\,
\barb{\chi}_\alpha(0)\quad .
\end{equation}
While being antisymmetric at $t=0$, $\barb{\chi}_\alpha
(0) \equiv \bbarb{A}\, \barb{\chi}_\alpha(0)$, such
eigenvectors generally develop a finite symmetric part
for $t\ne 0$. A Taylor expansion of the eigenvalue
equation for $\barb{\chi}_\alpha(t)$  in small $t$
reveals that a symmetric component appears in first
order (see Appendix~\ref{app:transPerturb} for details),
\begin{align}\label{eq:transSymmLead}
\barb{\chi}_\alpha(t) &= \left( \bbarb{1} - \frac{i\,
t}{\hbar\, n_\alpha(0)}\,\,\bbarb{\tau}\right)\bbarb{A}\,
\barb{\chi}_\alpha(t) + \mathcal{O}(t^2)\quad .
\end{align}
The leading-order \emph{transformer} matrix
$\bbarb{\tau}$ transforms the antisymmetric eigenvector
into the symmetric subspace. It is defined by
\begin{align}\label{eq:TauZerothOrder}
\bbarb{\tau} &=  i \,\hbar\,\, \bbarb{S}\,
\bbarb{\rho}^\prime(0)\, \bbarb{A} \quad ,
\end{align}
and its matrix elements are given by
Eq.~(\ref{eq:TransMat}). As is evident from
Eq.~(\ref{eq:transSymmLead}), the part $\bbarb{S}\,
\barb{\chi}_\alpha(t)$ of the eigenvector
$\barb{\chi}_\alpha(t)$ that is symmetric under index
exchange has a leading $\mathcal{O}(t)$ dependence on
relative time $t$. Thus, symmetric-pairing correlations
that are necessarily odd-in-$t$ to leading order are a
feature of any system for which the transformer
$\bbarb{\tau}$ with matrix elements given in
Eq.~(\ref{eq:TransMat}) is finite. 

As shown in Appendix~\ref{app:genTrans}, the transformer
scenario can be generalized to all orders in $t$. The
symmetric part of the eigenvector $\barb{\chi}_\alpha
(t)$ may be written as
\begin{align}\label{eq:TauDefImpl}
\bbarb{S}\,\barb{\chi}_\alpha(t) = \frac{-i\, t}{\hbar\,
n_\alpha(t)}\,\,\bbarb{\tau}_\alpha(t)\,\bbarb{A}\,
\barb{\chi}_\alpha(t) \quad ,
\end{align}
with the generalized transformer $\bbarb{\tau}_\alpha
(t)$ for the $\alpha$th eigenstate that is given
formally by
\begin{align}\label{eq:TauDefExpl}
\bbarb{\tau}_\alpha(t) &= \left[\,\bbarb{1} -
\frac{1}{n_\alpha(t)}\,\bbarb{S}\,\bbarb{\rho}(t)\,
\bbarb{S}\,\right]^{-1}\,\frac{i\,\hbar}{t}\,\bbarb{S}\,
\bbarb{\rho}(t)\,\bbarb{A} \quad .
\end{align}
Expanding these equations to leading order in $t$
readily yields Eqs.~\eqref{eq:transSymmLead} and
\eqref{eq:TauZerothOrder}.

\subsubsection{Generator mechanism}
\label{subsubsec:GenMech}

The transformer mechanism discussed above is not the
only way how symmetric pair correlations can arise. In
addition, we have to consider eigenvectors
$\barb{\chi}_\beta(t)$  with vanishing eigenvalues at
zero time; $n_\beta(0) = 0$. To make this explicit, we
rewrite the spectral decomposition
(\ref{eq:Rho(t)SpecDecomp}) of the T2bCM as follows:
\begin{align}\label{eq:newRho(t)Spec}
\bbarb{\rho}(t) = \sum_\alpha^\prime n_\alpha(t)\,
\barb{\chi}_\alpha(t)\, \barb{\chi}^\dagger_\alpha(t) +
\sum_\beta^{\prime\prime} n_\beta(t)\, \barb{\chi}_\beta
(t)\, \barb{\chi}^\dagger_\beta(t) \quad .
\end{align}
Here $\sum_\alpha^\prime$ is the restricted sum over
eigenvectors that  have nonzero eigenvalues $n_\alpha(0)
>0$. They are antisymmetric at $t=0$ and may or may not
develop symmetric parts for $t\ne 0$, depending on the
structure of the transformer $\bbarb{\tau}_\alpha(t)$.
In contrast, $\sum_\beta^{\prime\prime}$ contains only
eigenvectors whose associated eigenvalues vanish at
$t=0$. Those eigenvalues grow quadratically as a
function of $t$; $n_\beta(t)\sim \mathcal{O}(t^2)$,
because $n_\beta(t)\ge 0$ for all times. The
eigenvectors $\barb{\chi}_\beta(t)$ are of unknown
symmetry but may have a symmetric component
$\bbarb{S}\,\barb{\chi}_\beta(t) \ne \barb{0}$. To
identify these specific contributions, we consider the
symmetric-pair-correlation \emph{generator}
\begin{subequations}
\begin{align}\label{eq:genSchi}
\bbarb{\gamma}(t) &= \hbar^2\, \sum_\beta^{\prime\prime} 
\frac{n_\beta(t)}{t^2}\,\,\bbarb{S}\,\barb{\chi}_\beta(t)
\,\big[ \bbarb{S}\,\barb{\chi}_\beta(t) \big]^\dag
\quad , \\ \label{eq:genGen} 
&= \frac{\hbar^2}{t^2}\, \bbarb{S}\, \bbarb{\rho}(t)\,
\bbarb{S} - \sum_\alpha^\prime \, \bbarb{\tau}_\alpha(t)
\, \frac{\bbarb{A}\, \barb{\chi}_\alpha(t) \left[
\bbarb{A}\, \barb{\chi}_\alpha(t)
\right]^\dagger}{n_\alpha (t)}\,
\bbarb{\tau}^\dagger_\alpha(t) \, ,
\end{align}
\end{subequations}
where the last line follows using
Eqs.~\eqref{eq:TauDefImpl} and \eqref{eq:newRho(t)Spec}.
By construction, $\bbarb{\gamma}(t)$ can only be finite
if the fully symmetric part $\bbarb{S}\,\bbarb{\rho}(t)\,
\bbarb{S}$ of the T2bCM is not entirely accounted for by
the transformer mechanism, i.e., the generator embodies
any and all symmetric-pairing correlations arising by
complementary mechanisms. Its $t=0$ limit is given by
\begin{subequations}\label{eqs:generator}
\begin{align}\label{eq:genSchiSmallT}
\bbarb{\gamma} &= \frac{\hbar^2}{2} \sum_\beta^{\prime
\prime} n^{\prime\prime}_\beta(0)\,\, \bbarb{S}\,
\barb{\chi}_\beta(0)\,\big[ \bbarb{S}\,\barb{\chi}_\beta
(0) \big]^\dag \quad , \\ \label{eq:gammaForm}
&= \frac{\hbar^2}{2}\, \bbarb{S}\, \bbarb{\rho}^{\prime
\prime}(0)\,\bbarb{S} - \bbarb{\tau}\, \Cproj \big[
\bbarb{\rho}(0)\big]^{-1} \Cproj\, \bbarb{\tau}^\dagger
\quad .
\end{align}
\end{subequations}
Here we introduced the projector $\Cproj =
\sum_\alpha^\prime \barb{\chi}_\alpha(0)\,
\barb{\chi}_\alpha^\dagger(0)$ onto the subspace where
the 2bRDM has no zero modes. See Fig.~\ref{fig:rhoTrans}
for a refinement of the T2bCM structure presented
earlier in Fig.~\ref{fig:blockRho}. The expression given
in Eq.~(\ref{eq:gammaForm}) enables straightforward
calculation of $\gamma_{\vek{i}\vek{j}, \vek{k}\vek{l}}$
after diagonalizing the 2bRDM, using also the
transformer matrix elements from
Eq.~(\ref{eq:TransMat}), and
\begin{align}\label{eq:SRhoppS}
\frac{\hbar^2}{2}\left[ \bbarb{S}\, \bbarb{\rho}^{\prime
\prime}(0)\,\bbarb{S}\right]_{\vek{i}\vek{j},\vek{k}
\vek{l}} &= \nonumber \\[0.1cm] &\hspace{-2cm}
\frac{1}{4} \,\Langle \big(c^\dagger_\vek{i}\, H\,
c^\dagger_\vek{j} + c^\dagger_\vek{j}\, H\,
c^\dagger_\vek{i} \big) \big(c_\vek{l}\, H\, c_\vek{k}
+ c_\vek{k}\, H\, c_\vek{l} \big) \Rangle \,\,\, .
\end{align}
Inspecting $\bbarb{\gamma}$ for a given system of
interest provides basic insights into the nature of its
generated symmetric-pairing correlations, i.e., those
arising beyond the transformer mechanism.

It can be shown that the $\chi_\beta(0)$ having $n_\beta
(0)=0$ are the solutions of the eigenvalue equation
\begin{align}\label{eq:genEigenZero}
& \left\{ \Dproj\,\bbarb{\rho}^{\prime\prime}(0)\,\Dproj
- 2\,\Dproj\,\bbarb{\rho}^{\prime}(0)\,\Cproj\big[
\bbarb{\rho}(0)\big]^{-1}\Cproj\,\bbarb{\rho}^\prime(0)\,
\Dproj \right\}\barb{\chi}_\beta(0) \nonumber \\ &
\hspace{5cm} =\, n_\beta^{\prime\prime}(0)\,
\barb{\chi}_\beta(0) \,\, ,
\end{align}
where $\Dproj \equiv \sum_\beta^{\prime\prime}
\barb{\chi}_\beta(0)\,\barb{\chi}^\dagger_\beta(0) =
\bbarb{1} - \Cproj$. (See Appendix~\ref{app:symmPert}
for details, and Fig.~\ref{fig:rhoTrans} for an
illustration of the subspaces projected on by $\Dproj$
and $\Cproj$.) As per Eq.~(\ref{eq:genSchiSmallT}),
finiteness of the generator matrix $\bbarb{\gamma}$ is an
indicator for one or more of the eigenvectors
$\barb{\chi}_\beta(0)$ to have a symmetric part
$\bbarb{S}\,\barb{\chi}_\beta(0)\ne\barb{0}$. In the
particular case where $\barb{\chi}_\beta(0)$ is fully
symmetric, i.e., $\barb{\chi}_\beta(0) = \bbarb{S}\,
\barb{\chi}_\beta(0)$, Eq.~(\ref{eq:genEigenZero})
becomes an eigenvalue equation for $\bbarb{\gamma}$
(see further discussion in Appendix~\ref{app:symmPert}),
\begin{align}\label{eq:GammaEigEq}
\bbarb{\gamma}\,\barb{\chi}_\beta(0) = \frac{\hbar^2}{2}
\, n_\beta^{\prime\prime}(0)\,\barb{\chi}_\beta(0)\quad .
\end{align}
Thus, T2bCM eigenvectors that are fully symmetric in the
$t=0$ limit become also eigenvectors of the generator
matrix $\bbarb{\gamma}$ in that same limit.

\subsection{Macroscopic symmetric-pairing order}
\label{sec:SymmOP}

Generalizing the conventional Penrose-Onsager-Yang
approach~\cite{Penrose1956,Yang1962}, we associate pair
condensation into a superfluid state with having a
single term in the T2bCM's spectral decomposition
(\ref{eq:Rho(t)SpecDecomp}) with a macroscopic
eigenvalue, $n_0(t)\sim \mathcal{O}(N)$, signifying that
the pair orbital $\barb{\chi}_0(t)$ has a macroscopic
occupation number. The physical properties of the system
are then dominated by the condensate order
parameter~\cite{Leggett2006}
\begin{equation}\label{eq:purelyASrho0}
\barb{\phi}(t) = \sqrt{n_0(t)}\, \barb{\chi}_0(t) \,\, ,
\end{equation}
as the quantity $\barb{\phi}(t)/\sqrt{N}$ remains finite
in the thermodynamic limit $N\to\infty$. Thus, to leading
order in large $N$, the T2bCM factorizes as expressed in
Eq.~(\ref{eq:T2bCMfact}), in generalization of the
analogous factorisation in Gorkov's time-ordered
four-point function~\cite{Gorkov1958}.

In the $t=0$ limit, our criterion for superfluid order
to exist, and the definition (\ref{eq:purelyASrho0}) of
the associated order parameter, recover the case of
conventional even-frequency pair condensation. Clearly,
finiteness of $\barb{\phi}(0)$ requires $n_0(0)>0$. In
this case, $\barb{\chi}_0(0)$ is a macroscopic
eigenvector of the 2bRDM and, thus, necessarily
antisymmetric. The order parameter then specializes to
the form $\barb{\phi}(0) = \sqrt{n_0(0)}\,\barb{\chi}_0
(0)$ that is familiar from the
particle-number-conserving description of ordinary
Cooper-pair condensates~\cite{Leggett2006}.

For $t\ne 0$, $\barb{\phi}(t)$ as defined in
Eq.~(\ref{eq:purelyASrho0}) will have a finite
symmetric part $\bbarb{S}\,\barb{\phi}(t)$ whenever
$\bbarb{S}\,\barb{\chi}_0(t)\ne \barb{0}$. However, to be
properly part of the macroscopic pairing order parameter,
finiteness of $\bbarb{S}\,\barb{\phi}(t)/\sqrt{N}$ as
$N\to\infty$ is required. As we now show, the two
mechanisms identified in Sec.~\ref{sec:tVsIndexEig} lead
to two possible scenarios for a system to develop a
macroscopic $\bbarb{S}\, \barb{\phi}(t)$, indicating the
presence of symmetric-pairing order. 

\textit{Finite and macroscopic $n_0(0)$: Transformer
scenario.}
As discussed above, in the case where $n_0(0)$ is
finite, the macroscopic eigenvector $\barb{\chi}_0(t)$ is
antisymmetric at $t=0$, and the system exhibits
conventional antisymmetric-pairing (even-frequency)
superfluid order. By virtue of the transformer mechanism,
the order parameter (\ref{eq:purelyASrho0}) generally
has a symmetric part $\bbarb{S}\, \barb{\phi}(t)$ whose
leading-order small-$t$ dependence is linear. The
conditions under which $\bbarb{S}\, \barb{\phi}(t)/
\sqrt{N}$ will also be finite in the thermodynamic limit
depend on specifics of the physical system. Assuming this
to be the case, we define $\barb{\phi}^\mathrm{(s)}(t)
\equiv \bbarb{S}\, \barb{\phi}(t)$ and find
\begin{align}\label{eq:symmOPtr}
\barb{\phi}^\mathrm{(s)}(t) = -i\,
\frac{t}{\hbar}\, \frac{1}{\sqrt{n_0(0)}}\, \bbarb{\tau}
\, \bbarb{A}\,\barb{\chi}_0(0) + \mathcal{O}(t^2) \,\, .
\end{align}
Thus, in the presence of conventional antisymmetric-pair
condensation signaled by a macroscopic eigenvalue $n_0(0)
\sim \mathcal{O}(N)$ with associated antisymmetric
eigenvector $\barb{\chi}_0(0)\equiv \bbarb{A}\,
\barb{\chi}_0(0)$, the symmetric-pairing order parameter
$\barb{\phi}^\mathrm{(s)}(t)$ defined according to
(\ref{eq:symmOPtr}) emerges alongside the conventional
superfluid order. The transformer scenario covers the
existing proposals for odd-frequency superconductivity
to occur in conjunction with conventional even-frequency
order~\cite{Bergeret2005,Eschrig2007,Tanaka2007,
Kusunose2012,Sothmann2014,Black-Schaffer2013,Asano2015,
Triola2020,Chakraborty2022inter}. We present an
illustrative example in Sec.~\ref{sec:ExamplesTransf}. 

\textit{Vanishing $n_0(0)$, macroscopic $n_0^{\prime
\prime}(0)$: Generator scenario.}
In the absence of conventional macroscopic pair
condensation at $t=0$, the generator mechanism of
Sec.~\ref{sec:tVsIndexEig} provides an avenue for a
macroscopic $n_0(t)\sim \mathcal{O}(N)$ to emerge at
$t\ne 0$ and to cause symmetric-pairing order. As a
prerequisite, Eq.~(\ref{eq:genEigenZero}) must have a
single macroscopic eigenvalue $n_0^{\prime\prime}(0)\sim
\mathcal{O}(N)$. The small-$t$ limit for the macroscopic
eigenvalue of the T2bCM is then of the form $n_0(t)=t^2\,
n_0^{\prime\prime}(0)/2 + \mathcal{O}(t^3)$. In this
situation, the order parameter (\ref{eq:purelyASrho0})
has a leading-order small-$t$ dependence $\sim t$, and
its symmetric component
\begin{align}\label{eq:symmOPge}
\barb{\phi}^\mathrm{(s)}(t) = \frac{t}{\sqrt{2}}\,
\sqrt{n_0^{\prime\prime}(0)}\, \bbarb{S}\, \barb{\chi}_0
(0) + \mathcal{O}(t^2)
\end{align}
represents symmetric-pairing order. Note that the
symmetric order parameter $\barb{\phi}^\mathrm{(s)}(t)$
can be obtained from the dominant eigenpair of the
generator matrix $\bbarb{\gamma}(t)$ in the special case
where the eigenvector $\barb{\chi}_0(t)$ is fully
symmetric. More generally, a macroscopic eigenvalue of
the generator matrix implies a macroscopic symmetric
order parameter $\barb{\phi}^\mathrm{(s)}(t)$ under weak
additional assumptions. For a detailed discussion and
proof, see Appendix~\ref{app:genEvT2bCMEv}.

The generator scenario applies to instances of bulk
odd-frequency order where fermion-pair correlations exist
only at $t\ne 0$~\cite{Berezinskii1974,Kirkpatrick1991,
Balatsky1992,Abrahams1993,Emery1992,Balatsky1993,
Coleman1994,Coleman1995,Balatsky1994,Schrieffer1994,
Abrahams1995,Dahal2009,Fuseya2003,Kusunose2011a}. An
example of such a case is discussed in
Sec.~\ref{sec:ExampleGen}. Having a macroscopic
eigenvalue $n_0^{\prime\prime}(0)$ arising from
Eq.~(\ref{eq:genEigenZero}) [or, if applicable,
Eq.~(\ref{eq:GammaEigEq})] generally implies the
existence of some type of hidden order~\cite{Aeppli2020}
in the system of interest, and the generator scenario
describes how symmetric-pairing order arises as its
manifestation.
The order parameter $\barb{\phi}(t) \propto t\,
\sqrt{n_0^{\prime\prime}(0)/2}$ in this case and, thus,
the fermion-pair condensate fraction scales as $(t/
t^*)^2$ for $t < t^* \equiv \sqrt{2/n_0^{\prime\prime}
(0)}$. While this reduction may create practical
difficulties for revealing symmetric-pairing
order~\cite{Kornich2021} for too-small values of $t/
t^*$, the T2bCM is nonetheless dominated by the single
macroscopic contribution as per Eq.~\eqref{eq:T2bCMfact}
and, thus, exhibits the hallmarks of macroscopic quantum
coherence~\cite{Yang1962,Leggett2006}. We leave a more
detailed discussion of physical consequences associated
with the time scale $t^*$ and a condensate fraction
$\propto (t/t^*)^2$ for future research. In this context,
it is important to keep in mind that the time $t=t_1-t_2$
in Eq.~\eqref{eq:T2bCMgen} describes the internal
dynamics of the Cooper pair, akin to the time argument
in the dynamical Eliashberg
formalism~\cite{Eliashberg1960,Eliashberg1961} of
superconductivity, and not the out-of-equilibrium
dynamics that enters, e.g., a formulation in terms of a
time-dependent Ginzburg-Landau theory. The former is the
relative time $t$ of the two fermions forming the pair,
while the latter is the total time $(t_1+t_2)
/2$~\cite{Inkof2022}.

\textit{Higher-order scenario}.
In the transformer and generator scenarios described so
far, macroscopic pairing order manifests already in the
leading-order-in-$t$ contribution to $n_0(t)$. For the
transformer scenario, this is $n_0(0)$, the dominant
eigenvalue of the 2bRDM as per
Eq.~(\ref{eq:2bRDMeigen}). In the generator scenario
where $n_0(0)=0$, the leading term $n_0''(0)\, t^2/2$ in
the small-$t$ limit is macroscopic because $n_0''(0)$ is
the single macroscopic eigenvalue of the generator
matrix $\bbarb{\gamma}$ [or, more generally,
Eq.~(\ref{eq:genEigenZero})]. Other scenarios are
possible, where the macroscopic nature (i.e., scaling
with particle number $\sim N$) develops solely at finite
$t$ and becomes apparent only at higher orders in the
small-$t$ expansion of $n_0(t)$.

Our formalism describes odd-frequency superfluidity in
terms of the symmetric-pairing order parameter
$\barb{\phi}^\mathrm{(s)}(t)$, as defined in
Eqs.~(\ref{eq:symmOPtr}) and (\ref{eq:symmOPge}) for the
transformer scenario and the generator scenario,
respectively. How $\barb{\phi}^\mathrm{(s)}(t)$ can be
related to order parameters from number-nonconserving
theories is discussed in Appendix~\ref{app:relateOPs}.

\section{Meissner effect}\label{sec:Meissner}

One of the most striking applications of the
Penrose-Onsager-Yang formalism for conventional
superfluids is the possibility to derive the Meissner
effect and flux quantization solely from macroscopic
quantum coherence in the 2bRDM, without strong
assumptions about system properties~\cite{Sewell1990,
Nieh1995,Au1995,Sewell1997,Rampp2022}. Here we show how
the analogous description for the T2bCM developed in
Sec.~\ref{sec:Formalism} similarly lends itself to
discussing the Meissner effect for odd-frequency
superconductors, which has been a hotly debated
issue~\cite{Abrahams1995,Heid1995,Solenov2009,
Kusunose2011,Fominov2015,Linder2019,Parhizgar2021}. 

The general approach is based on the observation that a
spatial translation in a uniform and time-independent
magnetic field $\vek{B}$ is equivalent to a gauge
transformation of the magnetic vector potential $\vek{A}
(\rr)$~\cite{Levinson1970}. Specifically, for the
symmetric gauge $\vek{A}(\rr) = \frac{1}{2}\,\vek{B}
\times\rr$, it is straightforward to show the relation
\begin{subequations}\label{eqs:transGauge}
\begin{align}
\vek{A}(\rr + \vek{a}) &= \vek{A}(\rr) + \vek{\nabla}
\Lambda_{\vek{a}}(\rr) \quad , \\[0.2cm]
\Lambda_\vek{a}(\rr) &= -\vek{a}\cdot \vek{A}(\rr)\quad .
\end{align}
\end{subequations}
Gauge invariance of the overall system dynamics then
implies
\begin{equation}\label{eq:fermionGauge}
c_{\rr+\vek{a}} = \exp\left[ i\,\frac{q}{\hbar}\,
\Lambda_\vek{a}(\rr) \right]\, c_\rr \quad ,
\end{equation}
where $c_\rr$ denotes the annihilation operator for a
fermion with charge $q$ in real-space representation.
The relation (\ref{eq:fermionGauge}) can be used to infer
the transformational behavior of the 2bRDM
\begin{equation}
\rho_{\rr_i\,\rr_j ,\,\rr_k\,\rr_l} = \left\langle
c_{\rr_i}^\dagger\, c_{\rr_j}^\dagger\, c_{\rr_l}\,
c_{\rr_k} \right\rangle
\end{equation}
under translations. Together with the approximate
factorization (\ref{eq:RhoFact}) that represents
off-diagonal long-range order (ODLRO) in an ordinary
Cooper-pair condensate~\cite{Yang1962}, this leads to
the condition $\vek{B}=\vek{0}$, embodying the familiar
diamagnetic Meissner effect~\cite{Sewell1990,Nieh1995,
Au1995,Sewell1997,Rampp2022}.

Here we are interested in discussing the implications of
macroscopic coherence in the real-space representation
of the T2bCM
\begin{equation}
\rho_{\rr_i\,\rr_j ,\,\rr_k\,\rr_l}(t) = \left\langle
c_{\rr_i}^\dagger(t)\, c_{\rr_j}^\dagger\, c_{\rr_l}\,
c_{\rr_k}(t) \right\rangle
\end{equation}
instead of the 2bRDM $\rho_{\rr_i\,\rr_j ,\,\rr_k\,
\rr_l} \equiv \rho_{\rr_i\,\rr_j ,\,\rr_k\,\rr_l}(0)$.
Due to the gauge invariance of the system Hamiltonian
$H$ and, therefore, of the time-evolution operator $\exp
\big( \tfrac{-i\, t}{\hbar}\, H \big)$, the relation
(\ref{eq:fermionGauge}) generalizes to
\begin{equation}\label{eq:GaugeFinT}
c_{\rr+\vek{a}}(t) = \exp\left[ i\,\frac{q}{\hbar}\,
\Lambda_\vek{a}(\rr) \right]\, c_\rr(t) \quad .
\end{equation}
As a result, the T2bCM satisfies 
\begin{widetext}
\begin{equation}\label{eq:T2bCMtrans}
\rho_{\rr_i+\vek{a}\, \rr_j+\vek{a}\,, \,\rr_k+\vek{a}\,
\rr_l + \vek{a}}(t) = \exp\left\{ -i\,\frac{q}{\hbar}
\left[ \Lambda_\vek{a}(\rr_i) + \Lambda_\vek{a}(\rr_j)
- \Lambda_\vek{a}(\rr_k)  -\Lambda_\vek{a}(\rr_l)\right]
\right\}\, \rho_{\rr_i\,\rr_j ,\,\rr_k\,\rr_l}(t) \quad .
\end{equation}
\end{widetext}
The approximate factorization \eqref{eq:T2bCMfact} of the
T2bCM due to the existence of a macroscopic eigenvalue 
implies that the T2bCM is dominated by the
macroscopically coherent contribution at large length
scales. Specifically,
\begin{equation}\label{eq:ODLROt}
\rho_{\rr_i\,\rr_j ,\,\rr_k\,\rr_l}(t) \to \phi_{\rr_i\,
\rr_j}(t)\, \phi^\ast_{\rr_k\,\rr_l}(t)
\end{equation}
is valid for
\begin{equation}\label{eq:ODLROlength}
|\rr_i + \rr_j - \rr_k - \rr_l|/2 \ge L \gg |\rr_i -
\rr_j|\approx |\rr_k - \rr_l| \,\, .
\end{equation}
This is the generalization of the concept of
ODLRO~\cite{Yang1962} to the situation where $t\ne 0$,
with $L$ denoting the length scale beyond which the part
$\tilde{\rho}_{\vek{i} \vek{j},\vek{k}\vek{l}}(t)$ in
Eq.~\eqref{eq:T2bCMfact} has decayed and only the
contribution from the macroscopic eigenvector of the
T2bCM remains. Let us now assume that the
factorization \eqref{eq:ODLROt} is valid at a particular
value of the relative time $t$. Consistency of
(\ref{eq:ODLROt}) with (\ref{eq:T2bCMtrans}) then
requires that the order parameter transforms under
translations as
\begin{align}
& \phi_{\rr_i+\vek{a}\,\rr_j+\vek{a}}(t) = \nonumber
\\[0.1cm] & \hspace{1cm} f_\vek{a}\, \exp\left\{ -i\,
\frac{q}{\hbar}\, \left[ \Lambda_\vek{a}(\rr_i) +
\Lambda_\vek{a}(\rr_j)\right]\right\}\, \phi_{\rr_i\,
\rr_j}(t) \,\, ,
\end{align}
with a displacement-dependent phase factor $f_\vek{a}$.
Performing two consecutive translations, first along
vector $\vek{a}$ then along vector $\vek{b}$, leads to
the order-parameter transformation
\begin{widetext}
\begin{subequations}
\begin{align}
\phi_{\rr_i+\vek{a}+\vek{b}\,\rr_j+\vek{a}+\vek{b}}(t)&=
f_\vek{b}\,f_\vek{a}\,\exp\left\{-i\, \frac{q}{\hbar}\,
\left[ \Lambda_\vek{b}(\rr_i+\vek{a}) + \Lambda_\vek{b}
(\rr_j+\vek{a})+ \Lambda_\vek{a}(\rr_i) + \Lambda_\vek{a}
(\rr_j)\right]\right\}\,\phi_{\rr_i\, \rr_j}(t) \quad ,
\\[0.1cm] \label{eq:twoTransAB}
&= \exp\left[ i\, \frac{q}{\hbar}\,\vek{B}\cdot (\vek{a}
\times\vek{b})\right]\, f_{\vek{a}+\vek{b}}\,\exp\left\{
-i\, \frac{q}{\hbar}\, \left[ \Lambda_{\vek{a}+\vek{b}}
(\rr_i) + \Lambda_{\vek{a}+\vek{b}}(\rr_j)\right]\right\}
\, \phi_{\rr_i\, \rr_j}(t) \quad .
\end{align}
\end{subequations}
To obtain (\ref{eq:twoTransAB}), we made use of the
identity $\Lambda_\vek{b}(\rr+\vek{a})+\Lambda_\vek{a}
(\rr) = \Lambda_{\vek{a}+\vek{b}}(\rr)- \frac{1}{2}\,
\vek{B}\cdot(\vek{a}\times\vek{b})$. Performing the two
translations in opposite order yields, however,
\begin{subequations}
\begin{align}
\phi_{\rr_i+\vek{a}+\vek{b}\,\rr_j+\vek{a}+\vek{b}}(t)&=
f_\vek{a}\,f_\vek{b}\,\exp\left\{-i\, \frac{q}{\hbar}\,
\left[ \Lambda_\vek{a}(\rr_i+\vek{b}) + \Lambda_\vek{a}
(\rr_j+\vek{b})+ \Lambda_\vek{b}(\rr_i) + \Lambda_\vek{b}
(\rr_j)\right]\right\}\,\phi_{\rr_i\, \rr_j}(t) \quad ,
\\[0.1cm] \label{eq:twoTransBA}
&= \exp\left[ -i\, \frac{q}{\hbar}\,\vek{B}\cdot (\vek{a}
\times\vek{b})\right]\, f_{\vek{a}+\vek{b}}\,\exp\left\{
-i\, \frac{q}{\hbar}\, \left[ \Lambda_{\vek{a}+\vek{b}}
(\rr_i) + \Lambda_{\vek{a}+\vek{b}}(\rr_j)\right]\right\}
\, \phi_{\rr_i\, \rr_j}(t) \quad .
\end{align}
\end{subequations}
\end{widetext}
The required consistency of the results
(\ref{eq:twoTransAB}) and (\ref{eq:twoTransBA}) leads to
the condition
\begin{equation}\label{eq:fluxTrans}
\frac{2\, q}{\hbar}\, \vek{B}\cdot (\vek{a}\times\vek{b})
= 2\pi s \quad,
\end{equation}
with integer $s$. Arbitrariness of the displacements
$\vek{a}$ and $\vek{b}$ appearing on the left-hand side
of Eq.~(\ref{eq:fluxTrans}) makes it impossible for the
requirement to be generally satisfied, except for
$\vek{B}=\vek{0}$.
For systems with discrete translation invariance,
the smallest field allowed would amount to placing a
flux quantum in the unit cell, often corresponding to
extremely large magnetic-field values.

In the derivation of the Meissner effect from
ODLRO for conventional superconductors~\cite{Sewell1990,
Nieh1995,Au1995,Sewell1997,Rampp2022}, the factorization
\eqref{eq:ODLROt} is used at $t=0$. However, as our
arguments show, it is sufficient if this factorization
is valid at any particular value of $t$, as the
conclusion of Eq.~\eqref{eq:fluxTrans} is independent of
time. Remarkably, the diamagnetic Meissner effect then
follows even for purely odd-frequency superconductors in
the generator scenario.

We have explicitly demonstrated the
incompatibility of a translationally invariant quantum
state exhibiting macroscopic pairing order in the T2bCM
with a homogeneous magnetic field, generalizing the line
of reasoning originally advanced by
Sewell~\cite{Sewell1990,Sewell1997}. However, our
analysis including the dependence on relative time
extends also to related applications of Sewell's
arguments, e.g., to discuss flux quantization in
multiply connected geometries \cite{Nieh1995}, the
incompatibility of magnetic fields that vary slowly in
space~\cite{Au1995}, and the existence of vortex
lattices~\cite{Rampp2022}.

The above derivation of the Meissner effect applies to
the system as a whole, described by the full order
parameter $\phi_{\rr_i\, \rr_j}(t)$. Within our
formalism, it is thus impossible to discuss individual
contributions to the Meissner response arising from the
antisymmetric-pairing and symmetric-pairing parts in the
transformer scenario~\cite{Parhizgar2021}. 

Our formalism establishes on very general grounds the
connection between macroscopic pairing order in
the T2bCM and the conventional diamagnetic Meissner
effect, extending the previous understanding
\cite{Sewell1990,Nieh1995,Au1995,Sewell1997,Rampp2022}
about conventional antisymmetric-pairing orders to
symmetric-pairing order, even in the case where no
ODLRO is present in the 2bRDM. These arguments extend
beyond the range of validity for particle-nonconserving
approaches and are valid even in cases that
cannot be described by approximate theories, or by
Gorkov's anomalous pair-correlation function
\eqref{eq:AnomCorr}. On the other hand, the formalism
cannot yield direct insight about whether the macroscopic
order constitutes the energetically stable phase of the
system~\cite{Rampp2022} and is thus unable to resolve
on its own the ongoing stability debate~\cite{Heid1995,
Solenov2009,Kusunose2011,Fominov2015}. But, as outlined
briefly in Sec.~\ref{sec:intro}, the particle-conserving
theory is still crucial for testing the validity of
basic arguments aimed at establishing thermodynamic
stability of odd-frequency paring order.

\section{Application to specific instances of
symmetric-pairing order}\label{sec:Examples}

We illustrate the strength of the particle-conserving
formalism developed in the previous Section by applying
it to particular physical realizations of
symmetric-pairing (i.e., odd-in-time; odd-frequency)
order. In Sec.~\ref{sec:ExamplesTransf}, the transformer
scenario underlying odd-frequency pair correlations
emerging in a spin-balanced \textit{s}-wave
Fermi superfluid with Zeeman splitting is discussed.
Section~\ref{sec:ExampleGen} examines the generator
scenario realized for a composite-boson condensate that
has been proposed~\cite{Emery1992,Balatsky1993,
Coleman1994,Coleman1995,Balatsky1994,Schrieffer1994,
Abrahams1995,Dahal2009} as a generic type of system
exhibiting odd-frequency superconductivity in the
absence of ordinary (i.e., antisymmetric-pairing)
superfluid order.

\subsection{Zeeman-spin-split Fermi superfluid: Example
of transformed antisymmetric-to-symmetric pairing}
\label{sec:ExamplesTransf}

Motivated by the emergence of odd-frequency
superconductivity in superconductor-ferromagnet hybrid
structures~\cite{Bergeret2005}, the
\textit{s}-wave Fermi superfluid subject to
Zeeman spin splitting has been studied as a model
system for bulk odd-frequency
order~\cite{Chakraborty2022inter}. This system also
constitutes a particular realization of a generic
multiband superconductor where odd-frequency pairs are
expected to exist~\cite{Black-Schaffer2013,Triola2020,
Asano2015}. Its many-particle Hamiltonian is given by
\begin{align}\label{eq:HfHff}
H &=\sum_{\vek{q} \sigma} \, \epsilon_{\vek{q}\sigma}\,
c^\dagger_{\vek{q}\sigma} c_{\vek{q}\sigma} - U
\sum_{\vek{q}\vek{q}^\prime} c^\dagger_{\vek{q}\uparrow}
c^\dagger_{-\vek{q}\downarrow} c_{-\vek{q}^\prime
\downarrow} c_{\vek{q}^\prime\uparrow} \quad ,
\end{align}
where $\vek{q}$ and $\sigma\in \{\uparrow,\downarrow\}$
denote quantum numbers of a fermion's linear momentum
and spin-$1/2$ degrees of freedom, respectively, and
$U>0$ is the strength of the spin-singlet
orbital-\textit{s}-wave pairing interaction. The
single-particle energy dispersion is given by
\begin{align}\label{eq:spDisp}
\epsilon_{\vek{q}\uparrow(\downarrow)} =
\epsilon_{|\vek{q}|}\, \substack{-\\(+)}\, h\quad ,
\end{align}
where $h$ is the Zeeman spin-splitting energy, and the
$\vek{q}$-dependence is assumed to be isotropic.

Explicit calculation of the transformer matrix elements
defined by Eq.~(\ref{eq:TransMat}), using the
Hamiltonian $H$ from Eq.~(\ref{eq:HfHff}) and
identifying general fermion-state indices $\vek{i}$ with
the combined momentum and spin quantum numbers
$\vek{q}_i$ and $\sigma_i$, yields
\begin{equation}\label{eq:TauSpInt}
\bbarb{\tau} = \bbarb{\tau}^{(\mathrm{sp})} +
\bbarb{\tau}^{(\mathrm{int})} \quad ,
\end{equation}
with the single-particle and interaction-related
contributions to the transformer given by
\begin{widetext}
\begin{subequations}
\begin{align}\label{eq:TauDiagGen}
\tau^{(\mathrm{sp})}_{\vek{q}_i\sigma_i \, \vek{q}_j
\sigma_j\, , \, \vek{q}_k\sigma_k \, \vek{q}_l\sigma_l}
&= \frac{1}{2} \left(\epsilon_{\vek{q}_j \sigma_j}
- \epsilon_{\vek{q}_i\sigma_i} \right)\, \rho_{\vek{q}_i
\sigma_i \, \vek{q}_j\sigma_j\, ,\, \vek{q}_k\sigma_k\,
\vek{q}_l\sigma_l}(0) \quad , \\[0.2cm] \label{eq:UtoTau}
\tau^{(\mathrm{int})}_{\vek{q}_i\sigma_i \, \vek{q}_j
\sigma_j\, , \, \vek{q}_k\sigma_k \, \vek{q}_l\sigma_l}
&= - \frac{U}{2} \sum_\vek{q} \Big\langle
c^\dagger_{\vek{q}\uparrow}\, c^\dagger_{-\vek{q}
\downarrow} \, \big( \varsigma_j\, c^\dagger_{\vek{q}_i
\sigma_i} \, c_{-\vek{q}_j\, \bar{\sigma}_j} +
\varsigma_i\, c^\dagger_{\vek{q}_j\sigma_j}\,
c_{-\vek{q}_i\, \bar{\sigma}_i} \big) \, c_{\vek{q}_l
\sigma_l}\, c_{\vek{q}_k \sigma_k}\Big\rangle \quad .
\end{align}
\end{subequations}
\end{widetext}
For the compact notation of Eq.~(\ref{eq:UtoTau}), we use
$\bar{\sigma}$ to denote the opposite of $\sigma$; i.e.,
$\bar{\sigma} =\,\, \downarrow (\uparrow)$ if $\sigma =\,
\,\uparrow (\downarrow)$, and $\varsigma$ takes the
values $+1\, (-1)$ when $\sigma =\,\,\uparrow
(\downarrow)$. The structure of the expressions from
Eqs.~(\ref{eq:TauDiagGen}) and (\ref{eq:UtoTau})
suggests that the transformer is generally finite for a
system described by the Hamiltonian (\ref{eq:HfHff}),
implying that symmetric-pairing correlations exist.
However, for actual symmetric-pairing \emph{order} to
emerge, these correlations must become a macroscopic
property of the system so that $\barb{\phi}^\mathrm{(s)}
(t)$ given by Eq.~(\ref{eq:symmOPtr}) satisfies
$\barb{\phi}^\mathrm{(s)}(t)\sim\mathcal{O}(\sqrt{N})$
in the large-$N$ limit. As we now show, the
macroscopicity of symmetric-pairing correlations derives
from the antisymmetric-paring order in the superfluid.

In the absence of a Zeeman term (i.e., $h=0$),
the ground state of a fermion system with Hamiltonian
\eqref{eq:HfHff} is known to be a condensate of
\textit{s}-wave spin-singlet Cooper pairs. For finite
values of $h$, this may remain the case, as long as the
Zeeman splitting $h$ is smaller than a critical value
(the Chandrasekhar-Clogston
limit~\cite{Chandrasekhar1962,Clogston1962}), or
at arbitrary values of $h$ in the absence of
spin-relaxation processes where the populations of
spin-$\uparrow$ and spin-$\downarrow$ particles are
separately conserved and adjusted to be equal, as is
typical for ultracold-atom experiments~\cite{Chin2004,
Zwierlein2006,Schunck2006,Veeravalli2008}. In either
case, the zero-temperature ground state of the
Zeeman-spin-split Fermi gas remains unpolarized
with all macroscopic properties unchanged from the
superfluid phase at $h=0$~\cite{Chevy2010,
Radzihovsky2010,Cichy2014,Spada2021}. This superfluid
state thus has a conventional Cooper pair
condensate~\cite{Yang1962} with a 2bRDM with a single
macroscopic eigenvector $\barb{\chi}_0$ corresponding to
spin-singlet \textit{s}-wave pairing. Specifically, the
2bRDM factorizes to leading order in $N$ [see
Eq.~\eqref{eq:RhoFact}],
\begin{subequations}
\begin{align}
& \rho_{\vek{q}_i \sigma_i \, \vek{q}_j \sigma_j\, ,\,
\vek{q}_k\sigma_k\, \vek{q}_l\sigma_l} = \nonumber \\
&\hspace{0.2cm} n_0\, \chi_{0, \vek{q}_i\sigma_i\,
\vek{q}_j \sigma_j}\, \chi^{\ast}_{0, \vek{q}_k
\sigma_k \, \vek{q}_l\sigma_l} + \tilde{\rho}_{\vek{q}_i
\sigma_i \, \vek{q}_j \sigma_j\, ,\, \vek{q}_k\sigma_k\,
\vek{q}_l\sigma_l}\,\, ,
\end{align}
with $n_0\sim\mathcal{O}(N)$, and
\begin{align}\label{eq:asymmEV}
& \chi_{0, \vek{q}_i\sigma_i\, \vek{q}_j \sigma_j} =
\nonumber\\ & \hspace{0.4cm} \delta_{\vek{q}_j,
-\vek{q}_i}\left( \chi_{0, \vek{q}_i}\,\delta_{\sigma_i,
\uparrow}\,\, \delta_{\sigma_j ,\downarrow} -\, \chi_{0,
-\vek{q}_i}\, \delta_{\sigma_i , \downarrow}\,\,
\delta_{\sigma_j , \uparrow} \right) \,\, .
\end{align}
\end{subequations}
Substituting the expression (\ref{eq:TauDiagGen}) of
$\bbarb{\tau}^{(\mathrm{sp})}$ for $\bbarb{\tau}$ in the
formula (\ref{eq:symmOPtr}) of the symmetric-pairing
order parameter, using also (\ref{eq:asymmEV}) for the
macroscopic eigenvector $\barb{\chi}_0$, we find to
leading order in small $t$
\begin{align}\label{eq:SPhiZeem}
& \phi^\mathrm{(s)}_{\,\vek{q}_i\sigma_i\,\vek{q}_j
\sigma_j}(t)  = -i\,\, \frac{t}{\hbar}\,\, h\,\,
\sqrt{n_0}\,\, \delta_{\vek{q}_j, -\vek{q}_i}\nonumber
\\[0.1cm] & \hspace{1.2cm} \times \left(
\chi_{0, \vek{q}_i}\, \delta_{\sigma_i ,\uparrow}\,\,
\delta_{\sigma_j ,\downarrow} +\, \chi_{0, -\vek{q}_i}\,
\delta_{\sigma_i, \downarrow}\,\, \delta_{\sigma_j ,
\uparrow} \right)\,\, .
\end{align}
Thus, for finite Zeeman splitting $h$, odd-frequency
spin-triplet \textit{s}-wave order emerges alongside
even-frequency spin-singlet \textit{s}-wave superfluidity
in a Fermi gas~\cite{Chakraborty2022inter}.

In the derivation of (\ref{eq:SPhiZeem}), only the
single-particle contribution
$\bbarb{\tau}^{(\mathrm{sp})}$ to the transformer was
included, and this yielded the form of symmetric-pairing
order consistent with previous studies of the
Zeeman-spin-split Fermi
superfluid~\cite{Chakraborty2022inter}. Intriguingly,
this result emerged as a direct consequence of the Zeeman
splitting in the single-particle term of the Hamiltonian
(\ref{eq:HfHff}), while the interaction term was only
implicitly relevant as the source of conventional
antisymmetric-pairing order in the Fermi superfluid. In
principle, an explicitly interaction-dependent
contribution to symmetric-pairing order may arise based
on the part $\bbarb{\tau}^{(\mathrm{int})}$ of the
transformer, adding to the Zeeman-splitting-facilitated
portion (\ref{eq:SPhiZeem}). While this possibility
cannot be ruled out completely, we expect it to be
rarely relevant. See Appendix~\ref{app:UtauTriv}
for a more detailed discussion.

\subsection{Composite-boson condensate: Example of
generated symmetric-pairing order}
\label{sec:ExampleGen}

Early studies of odd-frequency superfluidity envisioned
situations where it appears on its own, i.e., not
alongside ordinary antisymmetric (even-frequency) Cooper
pairing of fermions as in the transformer scenario
discussed in the previous section. Nevertheless, the
possibility of another type of even-frequency order
being the fundamental origin of odd-frequency pair
correlations has tantalized ongoing research
efforts~\cite{Linder2019}. In particular, a relationship
between symmetric-pairing (odd-frequency) order and
composite-boson condensation has been actively
investigated~\cite{Emery1992,Balatsky1993,Coleman1994,
Coleman1995,Balatsky1994,Schrieffer1994,Abrahams1995,
Dahal2009}.

Here we consider a one-dimensional-lattice
realization of the model proposed in
Refs.~\cite{Balatsky1994,Schrieffer1994,Abrahams1995,
Dahal2009}. A system of itinerant electrons coupled to
bosonic spin excitations (magnons) is described by the
Hamiltonian
\begin{align}\label{eq:HfHbHffb}
H &= -K\,\sum_{r\sigma} \left( c^\dag_{r+1\,\sigma}
c_{r\sigma} + c^\dag_{r\sigma} c_{r+1\,\sigma} \right)
\nonumber \\ & \hspace{2cm} +\, J \sum_r \left(
c^\dag_{r\uparrow} c_{r\downarrow}\, b_r +
c^\dag_{r\downarrow} c_{r\uparrow}\, b^\dagger_r
\right) \,\, ,
\end{align}
where $c^\dag_{r\sigma}$ creates an electron with spin
$\sigma$ at lattice position $r$, and $K > 0$ is the
nearest-neighbour electron-hopping energy. Creation and
annihilation of a magnon at site $r$ [described by boson
operators $b^\dagger_r$ and $b_r$, respectively] incurs
a spin flip of electrons at the same site, with an
associated (exchange-)energy scale $J$. In the
following, we assume that the ground state used to
calculate the expectation values determining the 2bRDM
[Eq.~(\ref{eq:2bRDMdef})], the transformer
$\bbarb{\tau}$ [Eq.~(\ref{eq:TransMat})] and
the generator $\bbarb{\gamma}$ [Eq.~(\ref{eq:gammaForm})
with (\ref{eq:SRhoppS})] exhibits no independent
electron-pair or magnon condensates. This implies that
the 2bRDM has no macroscopic eigenvalue.

Without a macroscopic eigenvalue of the 2bRDM, there is
no possibility for symmetric-pairing order to arise via
the transformer scenario. \emph{Generated}
symmetric-pairing order, on the other hand, emerges when
the generator $\bbarb{\gamma}$ has an eigenvector with
macroscopic eigenvalue. See Eq.~(\ref{eq:GammaEigEq})
[more generally, Eq.~(\ref{eq:genEigenZero}) and further
discussion in Appendix~\ref{app:genEvT2bCMEv}].
Calculation of the generator, defined in
Eq.~(\ref{eq:gammaForm}), still requires knowledge of
the system's transformer matrix $\bbarb{\tau}$. We
present details of how $\bbarb{\tau}$ and
$\bbarb{\gamma}$ are obtained in
Appendix~\ref{app:elMagTauGen}. With the assumption that
the system ground state is an eigenstate of both the
electron number operator $\hat{N}\equiv \sum_{r \sigma}
c^\dagger_{r\sigma} c_{r\sigma}$ and the magnon number
operator $\hat{N}_\mathrm{b}\equiv \sum_r b^\dagger_r
b_r$, the generator matrix for the system described by
the Hamiltonian (\ref{eq:HfHbHffb}) is found to only
have terms dependent on the exchange-coupling strength
$J$;
\begin{widetext}
\begin{align}\label{eq:Gamma1DLatt}
\gamma_{r_i\sigma_i\, r_j\sigma_j\, ,\, r_k\sigma_k\,
r_l\sigma_l} = \frac{J^2}{4} &\left[ \,\delta_{\sigma_i,
\uparrow}\,\delta_{\sigma_j,\uparrow}\,\delta_{\sigma_k,
\uparrow}\,\delta_{\sigma_l,\uparrow}\,\left\langle\,
\Big( c^\dag_{r_i\uparrow}\, c^\dag_{r_j\downarrow}\,
b^\dag_{r_j} - c^\dag_{r_i\downarrow}\, c^\dag_{r_j
\uparrow}\, b^\dag_{r_i} \Big)\,\Big( b_{r_l}\, c_{r_l
\downarrow}\, c_{r_k\uparrow} - b_{r_k}\, c_{r_l
\uparrow}\, c_{r_k\downarrow}\Big)\,\right\rangle\right.
\nonumber\\[4pt] &\hspace{0.05cm} +\, \delta_{\sigma_i,
\downarrow}\,\delta_{\sigma_j,\downarrow}\,
\delta_{\sigma_k,\downarrow}\,\delta_{\sigma_l,
\downarrow}\,\left\langle\,\Big( c^\dag_{r_i\uparrow}\,
c^\dag_{r_j\downarrow}\, b_{r_i} - c^\dag_{r_i
\downarrow}\, c^\dag_{r_j\uparrow}\, b_{r_j}\Big)\,
\Big( b^\dag_{r_k}\, c_{r_l\downarrow}\, c_{r_k\uparrow}
- b^\dag_{r_l}\, c_{r_l\uparrow}\, c_{r_k\downarrow}\Big)
\, \right\rangle \nonumber\\[4pt] &\hspace{0.05cm} +\,
\left\langle c^\dag_{r_i\uparrow}\, c^\dag_{r_j\uparrow}
\,\Big(\delta_{\sigma_i,\uparrow}\,\delta_{\sigma_j,
\downarrow}\,b_{r_j} -\delta_{\sigma_i,\downarrow}\,
\delta_{\sigma_j,\uparrow}\, b_{r_i} \Big)\, \Big(
\delta_{\sigma_k,\uparrow}\,\delta_{\sigma_l,\downarrow}
\, b^\dag_{r_l} -\delta_{\sigma_k,\downarrow}\,
\delta_{\sigma_l,\uparrow} \, b^\dag_{r_k} \Big)\, c_{r_l
\uparrow}\, c_{r_k\uparrow} \right\rangle \nonumber
\\[4pt] &\hspace{0.05cm}+\,\hspace{-0.048cm} \left.
\left\langle c^\dag_{r_i\downarrow}\, c^\dag_{r_j
\downarrow}\,\Big( \delta_{\sigma_i,\uparrow}\,
\delta_{\sigma_j,\downarrow}\, b^\dag_{r_i} -
\delta_{\sigma_i,\downarrow}\,\delta_{\sigma_j,\uparrow}
\, b^\dag_{r_j}\Big)\,\Big( \delta_{\sigma_k,\uparrow}\,
\delta_{\sigma_l,\downarrow}\, b_{r_k} -
\delta_{\sigma_k,\downarrow}\,\delta_{\sigma_l,\uparrow}
\, b_{r_l} \Big)\, c_{r_l\downarrow}\, c_{r_k
\downarrow} \right\rangle\,\right] \, .
\end{align}
The right-hand side of Eq.~(\ref{eq:Gamma1DLatt})
contains various generalized three-body reduced density
matrices describing correlations between an
itinerant-electron pair and a magnon. A macroscopic
eigenvalue of $\bbarb{\gamma}$ would have to arise from
a hidden order~\cite{Linder2019,Aeppli2020} involving
such combinations of electronic and magnetic degrees of
freedom. Our particle-number-conserving formalism
enables a detailed discussion of the possibility that
condensation of bosonic fermion pairs coupled with
bosonic spin excitations underpin symmetric-pairing
order in the system under consideration.

The generator's connection with composite-boson
condensation is made particularly apparent by focusing
on its matrix elements satisfying $r_i = r_j = r$ and
$r_k = r_l = r'$,
\begin{align}\label{eq:GammaUpUpUpUp}
&\gamma_{r\sigma_i\, r\sigma_j\, ,\, r'\sigma_k\, r'
\sigma_l} = J^2\, \big[ \, \delta_{\sigma_i,\uparrow}\,
\delta_{\sigma_j,\uparrow}\,\delta_{\sigma_k,\uparrow}\,
\delta_{\sigma_l,\uparrow}\,\,\Langle c^\dag_{r\uparrow}
\, c^\dag_{r\downarrow}\, b^\dag_r\, b_{r'} \, c_{r'
\downarrow}\, c_{r'\uparrow} \Rangle + \delta_{\sigma_i,
\downarrow}\, \delta_{\sigma_j,\downarrow}\,
\delta_{\sigma_k,\downarrow}\,\delta_{\sigma_l,
\downarrow}\,\,\Langle c^\dag_{r\uparrow}\, c^\dag_{r
\downarrow}\, b_r\, b^\dag_{r'}\, c_{r'\downarrow}\,
c_{r'\uparrow} \Rangle\,\big] \,\, .
\end{align}
\end{widetext}
The combinations of electron-pair operators with spin
excitations appearing on the right-hand-side of
Eq.~(\ref{eq:GammaUpUpUpUp}) correspond to the
order-parameter structure of the composite-boson
condensate proposed, e.g., in Refs.~\cite{Abrahams1995,
Dahal2009}. Hypothesizing a form of the
fixed-particle-number ground state that maximises the
composite-boson condensate (generalising an approach
pioneered by Yang~\cite{Yang1962}), we find that the
three-body reduced density matrix with elements
\begin{align}\label{eq:3bRDM}
\rho^{\mathrm{(3b)}}_{r_i\sigma_i\, r_j\sigma_j\, r_o ,\,
r_k\sigma_k\, r_l\sigma_l \, r_p} = \Langle c^\dag_{r_i
\sigma_i}\, c^\dag_{r_j\sigma_j}\, b^\dag_{r_o}\,
b_{r_p} \, c_{r_l\sigma_l}\, c_{r_k\sigma_k} \Rangle
\end{align}
factorizes to leading order in $N$ (see
Appendix~\ref{app:compCondRDM})
\begin{subequations}
\begin{align}\label{eq:3bRDMfact}
&\rho^{\mathrm{(3b)}}_{r_i\sigma_i\, r_j\sigma_j\, r_o
,\, r_k\sigma_k\, r_l\sigma_l \, r_p} = \nonumber
\\[0.1cm] & n_0^{\mathrm{(3b)}}\, \chi_{0, r_i\sigma_i\,
r_j\sigma_j\, r_o}^{\mathrm{(3b)}}\, \chi_{0, r_k\sigma_k
\, r_l\sigma_l\, r_p}^{\mathrm{(3b)}\ast} +
\tilde{\rho}^{\mathrm{(3b)}}_{r_i\sigma_i\, r_j\sigma_j
\, r_o\, r_k\sigma_k\, r_l\sigma_l \, r_p}\, .
\end{align}
Here $n_0^{\mathrm{(3b)}}\sim \mathcal{O}(N)$, the
residual matrix $\bbarb{\tilde{\rho}}^{\mathrm{(3b)}}$
has no macroscopic contribution, and the macroscopic
eigenvector of the three-body reduced density matrix has
the form
\begin{align}\label{eq:3bRDMeigvec}
\chi_{0, r_i\sigma_i\, r_j\sigma_j\, r_o}^{\mathrm{(3b)}}
= \chi_0^\mathrm{(3b)} \, \delta_{r_j, r_i}\,
\delta_{r_o, r_i} \left( \delta_{\sigma_i, \uparrow}\,
\delta_{\sigma_j,\downarrow} - \delta_{\sigma_i,
\downarrow}\, \delta_{\sigma_j, \uparrow} \right) .
\end{align}
\end{subequations}
As a result, the generator (\ref{eq:Gamma1DLatt})
satisfies the eigenvalue equation (\ref{eq:GammaEigEq})
with a symmetric eigenvector $\barb{\chi}_0(0)$;
\begin{subequations}
\begin{align}
\chi_{0, r_i\sigma_i\, r_j\sigma_j}(0) &=
\chi_0^\mathrm{(3b)}\,\, \delta_{r_j, r_i}\,\,
\delta_{\sigma_i,\uparrow}\,\,\delta_{\sigma_j,\uparrow}
\,\,\,\, , \\
n_0^{\prime\prime}(0) &= \frac{2 J^2}{\hbar^2}\,
n_0^\mathrm{(3b)} \,\, .
\end{align}
\end{subequations}
The generated symmetric-pairing order parameter defined
in Eq.~(\ref{eq:symmOPge}) then takes the form
\begin{align}\label{eq:comBosSP}
\phi^\mathrm{(s)}_{\,r_i\sigma_i\, r_j\sigma_j}(t) =
\frac{t}{\hbar}\,\, J\,\, \sqrt{n_0^\mathrm{(3b)}} \,\,
\chi_0^\mathrm{(3b)}\,\, \delta_{r_j, r_i}\,\,
\delta_{\sigma_i,\uparrow}\,\, \delta_{\sigma_j,\uparrow}
\end{align}
to leading order in small $t$. Thus, odd-frequency
spin-polarized-triplet pairing of electrons emerges from
the condensation of \textit{s}-wave singlet electron
pairs coupled to magnons~\cite{Balatsky1994,
Schrieffer1994,Abrahams1995}.

The formalism illustrated here provides a general recipe
for systematically identifying avenues toward generating
symmetric-pairing order. Given a microscopic model, the
reduced density matrices making up the generator
designate the channels for equal-time many-particle
condensation that underpins symmetric-pairing order. In
the above consideration, we assumed condensation of
singlet-fermion pairs coupled to magnons and found
symmetric-pairing correlations in the spin-polarized
triplet channel. In this scenario, the relevant terms of
the generator (\ref{eq:Gamma1DLatt}) are those from the
first two lines. Alternatively, the structure of terms
in the last two lines of Eq.~(\ref{eq:Gamma1DLatt})
implies that composite-boson condensation involving
triplet fermion pairs would generate symmetric-singlet
pairing order.

Here we have established a direct causal link between the
presence of a composite-boson condensate and macroscopic
symmetric-pairing correlations. This example demonstrates
how generated symmetric-pairing order is generally a
consequence of some type of hidden multi-particle
condensation ensuring a macroscopic eigenvalue of the
generator matrix. The general form of the generator as
given in Eq.~(\ref{eq:gammaForm}), in conjunction with
the matrix elements (\ref{eq:SRhoppS}), should enable a
comprehensive classification of system Hamiltonians that
can give rise to generated odd-frequency superfluidity.

\section{Conclusions}\label{sec:Conc}

We present a formalism to describe odd-in-time (also
called \emph{odd-frequency}) pairing, i.e., pair
correlations that are only occurring between two fermions
present at different times $t_1$ and $t_2$. Our approach
is based on a thorough study of the time-dependent
two-body correlation matrix (T2bCM) $\bbarb{\rho}(t_1,
t_2)$ [Eq.~(\ref{eq:T2bCMgen})]. The T2bCM is
well-defined in real physical systems that conserve
particle number and could therefore, in principle, be
probed directly in experiments similar to those that
have recently been performed~\cite{Holten2022} or
proposed~\cite{Malas2023} to measure even-in-time
($t_1=t_2$) pair correlations. In addition, the T2bCM
has the required properties for being a suitable
generalization of the two-body reduced density matrix
(2bRDM) $\bbarb{\rho}\equiv\bbarb{\rho}(0,0)$ utilized
in particle-conserving descriptions of conventional
fermion superfluidity~\cite{Yang1962,Leggett2006}. While
the formalism applies more generally, we have focused on
equilibrium situations described by a time-independent
Hermitian Hamiltonian $H$ conserving fermion number $N$.
In that case, the T2bCM depends only on the relative
time $t = t_1 - t_2$; $\bbarb{\rho}(t_1, t_2) =
\bbarb{\rho}(t, 0)\equiv \bbarb{\rho}(t)$, and
$\bbarb{\rho}(0)$ corresponds to the 2bRDM. Generalizing
the particle-number-conserving description of
fermion-pair condensation~\cite{Penrose1956,Yang1962,
Leggett2006}, the existence of a pair condensate is
signalled by $\bbarb{\rho}(t)$ having an eigenvector
$\barb{\chi}_0(t)$ with macroscopic eigenvalue $n_0(t)
\sim \mathcal{O}(N)$. The condensate order parameter
$\phi(t)$ [Eq.~(\ref{eq:purelyASrho0})] then satisfies
$\phi(t)\sim \mathcal{O}(\sqrt{N})$ in the large-$N$
(i.e., the thermodynamic) limit. We establish odd-in-$t$
pairing order to be associated with the symmetric part
$\barb{\phi}^\mathrm{(s)}(t)$ of the order parameter,
i.e., the part that does not change sign under fermion
exchange. 
Our derivation of the diamagnetic Meissner effect
from off-diagonal long-range order at any value of $t$
suggests that the presence of a macroscopic eigenvalue
of the T2bCM implies superfluid phenomena, even if no
macroscopic pairing is present in the 2bRDM.
Two scenarios are identified for
symmetric-pairing order embodied by
$\barb{\phi}^\mathrm{(s)}(t)$ to exist.

The \emph{transformer} scenario can occur when $n_0(0)\ne
0$ is macroscopic and, thus, $\barb{\chi}_0(0)$ is an
antisymmetric eigenvector of the 2bRDM. In such a case,
$\barb{\phi}^\mathrm{(s)}(t)$ arises from the
transformation of even-in-$t$ order into odd-in-$t$
order, facilitated by the transformer matrix
$\bbarb{\tau}$ [Eq.~(\ref{eq:symmOPtr}) with
Eq.~(\ref{eq:TauZerothOrder})]. We illustrate the
transformer scenario using the spin-polarized Fermi
superfluid as an example
[Sec.~\ref{sec:ExamplesTransf}]. The transformer matrix
$\bbarb{\tau}$ provides a natural way to quantify the
propensity for symmetric-pairing (i.e., odd-frequency)
order to emerge in the presence of an even-frequency
Cooper-pair condensate. It is thus similar to the
recently proposed superconducting-fitness
measure~\cite{Ramires2016,Ramires2018} whose connection
with odd-frequency superconductivity has been
established within Bogoliubov-de$\,$Gennes mean-field
theory~\cite{Triola2020}. Our formalism could be
utilized for further detailed investigation of the
superconducting-fitness concept, especially its
generalization beyond mean-field theory.

The alternative to the transformer scenario is the
\emph{generator} scenario, where symmetric-pairing
order emerges without an ordinary Cooper-pair condensate
present. The order parameter $\barb{\phi}^\mathrm{(s)}
(t)$ is then associated with an eigenvector
$\barb{\chi}_0(t)$ of the T2bCM $\bbarb{\rho}(t)$ that
has a symmetric part at $t=0$ [Eq.~(\ref{eq:symmOPge})]
and satisfies the eigenvector equation
(\ref{eq:genEigenZero}) with a macroscopic eigenvalue
$n_0''(0)\sim\mathcal{O}(N)$. The propensity of a system
to host generated symmetric-pairing order is embodied in
the generator matrix $\bbarb{\gamma}$
[Eqs.~(\ref{eqs:generator})]. We elucidate how the
generator scenario transpires in a model system of
itinerant electrons coupled to magnons due to the
presence of a composite-boson (fermion pair + magnon)
condensate [Sec.~\ref{sec:ExampleGen}]. Our formalism
lends itself to establishing the direct link between a
system's hidden order~\cite{Aeppli2020} and
odd-frequency superconductivity. Future research could
perform a systematic search for realizations of
symmetric-pairing order in other many-particle model
systems~\cite{Coleman2015} based  on consideration of
their generator matrices. 
Exotic multi-particle condensates~\cite{Zhang2018,
Nazaryan2024} are particularly promising candidates for
unveiling generated odd-frequency order.

Extensions of the present theoretical description could
consider symmetric-pairing order associated with a
higher-order $t$ dependence. Further investigation of
the transformer and generator matrices would also inform
efforts to design systems with odd-frequency superfluid
order present. A search for measurable quantities whose
response functions are related to a system's generator
matrix may yield indirect experimental probes of
generated symmetric-pairing order. More generally, due
to its focus on physically realistic
particle-number-conserving quantities, we expect our
work to boost the overall development of direct
detection schemes for odd-in-$t$ pairing. In addition,
the fixed-$N$ theory developed here is ideally suited
to study fermion pairing occurring in cold-atom
gases~\cite{Holten2022,Malas2023} and nuclear
matter~\cite{Dean2003,Sedrakian2019,Yang2023}, which
are new platforms for realizing and investigating
symmetric-pairing order~\cite{Kalas2008,Arzamasovs2018,
Linder2019}.

\begin{acknowledgments}

The authors are grateful to Philip M.R.\ Brydon for
pointing out the relevance of superconducting fitness in
relation to odd-frequency pairing. K.T.\ also thanks
Alexander V.\ Balatsky for useful discussions. J.B.\
acknowledges support from the Marsden Fund of New
Zealand (contract no.\ MAU2007) from government funding
administered by the Royal Society Te Ap\=arangi.
J.S.\ was supported by the Deutsche
Forschungsgemeinschaft (DFG, German Research Foundation)
-- TRR 288-422213477 Elasto-Q-Mat, project B01.

\end{acknowledgments}

\appendix

\section{Derivation of general transformer}
\label{app:genTrans}

We start by applying the symmetrizer $\bbarb{S}$ on both
sides of the eigenvalue equation for $\barb{\chi}_\alpha
(t)$, which yields
\begin{align}\label{eq:symmPart}
n_\alpha(t)\, \bbarb{S}\, \barb{\chi}_\alpha(t) &=
\bbarb{S}\, \bbarb{\rho}(t)\, \barb{\chi}_\alpha(t)
\nonumber \\ &= \bbarb{S}\, \bbarb{\rho}(t)\, \left[
\bbarb{S}\, \barb{\chi}_\alpha(t) + \bbarb{A}\,
\barb{\chi}_\alpha(t) \right] \quad .
\end{align}    
Assuming a nonvanishing antisymmetric contribution
$\bbarb{A}\,\barb{\chi}_\alpha(t)$, the implicit
relation (\ref{eq:symmPart}) between an eigenvector's
symmetric and antisymmetric parts is formally resolved as
\begin{align}\label{eq:intermedTau}
\bbarb{S}\, \barb{\chi}_\alpha(t) &= \left[n_\alpha(t)\,
\bbarb{1} - \bbarb{S}\,\bbarb{\rho}(t)\, \bbarb{S}\,
\right]^{-1} \bbarb{S}\,\bbarb{\rho}(t)\, \bbarb{A}\,
\barb{\chi}_\alpha(t) \,\, .
\end{align}
Here the invertibility of the expression between the
square brackets requires $n_\alpha(t)$ to be distinct
from any eigenvalues of $\bbarb{S}\,\bbarb{\rho}(t)\,
\bbarb{S}$. This can be guaranteed for any $n_\alpha(0)>
0$ and small-enough $t$ because all eigenvalues of
$\bbarb{S} \,\bbarb{\rho}(t)\,\bbarb{S}$ are $\sim
\mathcal{O}(t^2)$ and, thus, can become degenerate with
$n_\alpha(t)$ only at strictly finite $t$. In the
physically relevant transformer scenario where
$n_\alpha(t)$ is the only macroscopic eigenvalue both at
zero and finite $t$, the invertibility of the bracketed
expression is guaranteed.

The derived expression \eqref{eq:intermedTau} resolves
to Eq.~(\ref{eq:TauDefImpl}) in the main text, with
Eq.~(\ref{eq:TauDefExpl}) defining the general form of
the transformer matrix $\bbarb{\tau}_\alpha(t)$ under
the additional assumption that the eigenvalue
$n_\alpha(t)$ does not vanish. The definition of the
transformer matrix $\bbarb{\tau}_\alpha(t)$ ensures that
it has an $\alpha$-independent zero-$t$ limit
$\bbarb{\tau}_\alpha(0) = \bbarb{\tau}$ as given in
Eq.~\eqref{eq:TauZerothOrder}, with matrix elements
given by Eq.~\eqref{eq:TransMat}.

\section{Small-\texorpdfstring{$\bm{t}$}{$t$} expansion
for T2bCM eigenvectors and eigenvalues}
\label{app:recurse}

We consider a general eigenvalue equation $\bbarb{\rho}
(t)\,\barb{\chi}_\alpha(t) = n_\alpha(t)\,
\barb{\chi}_\alpha(t)$ for the T2bCM $\bbarb{\rho}(t)$.
Inserting the expansion of $\bbarb{\rho}(t)$ upto
$\mathcal{O}(t^2)$ given in Eq.~(\ref{eq:RhoTimeExp})
from the main text, as well as analogous expansions for
the eigenvector $\barb{\chi}_\alpha(t)$ and eigenvalue
$n_\alpha(t)$,
\begin{subequations}
\begin{align}
\barb{\chi}_\alpha(t) &= \barb{\chi}_\alpha(0) + t\,
\barb{\chi}_\alpha^\prime(0) + \frac{t^2}{2}\,
\barb{\chi}_\alpha^{\prime\prime}(0) + \mathcal{O}(t^3)
\quad , \\ \label{eq:eigenVexp}
n_\alpha(t) &= n_\alpha(0) + t\, n_\alpha^\prime(0) +
\frac{t^2}{2}\, n_\alpha^{\prime\prime}(0) + \mathcal{O}
(t^3) \quad ,
\end{align}
\end{subequations}
and equating coefficients of powers $t^0$, $t^1$ and
$t^2$, yields the relations
\begin{subequations}\label{eq:Ot012terms}
\begin{align}\label{eq:Ot0term}
\left[ n_\alpha(0) - \bbarb{\rho}(0) \right] 
\barb{\chi}_\alpha(0) &= \barb{0} \quad , \\
\label{eq:Ot1term}
\left[ n_\alpha(0) - \bbarb{\rho}(0) \right]
\barb{\chi}_\alpha^\prime(0) &= \left[
\bbarb{\rho}^\prime(0) - n_\alpha^\prime(0) \right]
\barb{\chi}_\alpha(0) \,\, , \\ \label{eq:Ot2term}
\left[ n_\alpha(0) - \bbarb{\rho}(0) \right]
\barb{\chi}_\alpha^{\prime\prime}(0) &= \left[
\bbarb{\rho}^{\prime\prime}(0) - n_\alpha^{\prime\prime}
(0) \right] \barb{\chi}_\alpha(0) \nonumber \\ &
\hspace{0.5cm} + 2 \left[ \bbarb{\rho}^\prime(0) -
n_\alpha^\prime(0) \right]\barb{\chi}_\alpha^\prime(0)
\,\, .
\end{align}
\end{subequations}
Relations involving higher-order $t$-derivatives can be
straightforwardly obtained by expanding each quantity
upto a higher power of $t$, but any features we are
interested in as part of the present work can be readily
illustrated based on Eqs.~(\ref{eq:Ot0term}),
(\ref{eq:Ot1term}) and (\ref{eq:Ot2term}). The formal
structure of these relations is analogous to that
emerging in the context of time-independent perturbation
theory in quantum mechanics~\cite{Sakurai2011}. However,
in the situation focused on here, the perturbation is
controlled by the small parameter $t$. A further twist
on the familiar perturbation-theory approach is that we
are interested in separating symmetric and antisymmetric
contributions to the eigenvector $\barb{\chi}_\alpha(t)$.

Multiplication of Eq.~(\ref{eq:Ot0term}) from the left
with $\bbarb{S}$ and using $\bbarb{S}\, \bbarb{\rho}(0) =
\bbarb{0}$ yields
\begin{equation}\label{eq:t=0cases}
n_\alpha(0)\, \bbarb{S}\, \barb{\chi}_\alpha(0) =
\barb{0} \quad .
\end{equation}
Thus, in the $t=0$ limit, eigenvectors are purely
antisymmetric if the eigenvalue is nonzero, because
$n_\alpha(0)\ne 0$ implies $\bbarb{S}\,
\barb{\chi}_\alpha(0)=\barb{0}$ based on
Eq.~(\ref{eq:t=0cases}). As finite-$t$ corrections can
contain both antisymmetric and symmetric contributions,
$\barb{\chi}_\alpha(t)$ could remain purely
antisymmetric or become a mixture of antisymmetric and
symmetric parts. We discuss this case in
Appendix~\ref{app:transPerturb} below.

The case of $n_\alpha(0)=0$ has to be treated
carefully because of the potential for degeneracy between
symmetric and antisymmetric subspaces. The degeneracy
may be lifted at finite $t$, where mixing between the
two sectors may occur, which will affect the $t\to 0$
limit. We use degenerate pertubation theory to treat
this case in Appendix~\ref{app:symmPert}.

\subsection{Eigenvectors with finite eigenvalues at
\texorpdfstring{$\bm{t=0}$}{$t=0$}}
\label{app:transPerturb}

Equation~(\ref{eq:Ot0term}) constitutes the eigenvalue
equation for the 2bRDM $\bbarb{\rho}(0)$. We assume that
the eigenvalue problem for $\bbarb{\rho}(0)$ has been
solved for a particular system of interest. Here we focus
on the set of nonzero eigenvalues, $n_\alpha(0)>0$, and
their associated $\barb{\chi}_\alpha(0)$. As discussed
above, these are fully antisymmetric; $\barb{\chi}_\alpha
(0) \equiv \bbarb{A}\, \barb{\chi}_\alpha(0)$.

Multiplying Eq.~(\ref{eq:Ot1term}) with $\bbarb{S}$ from
the left and remembering $\bbarb{S}\, \bbarb{\rho}(0) =
\bbarb{0}$, as well as $\bbarb{S}\, \barb{\chi}_\alpha
(0) = \barb{0}$, we find
\begin{equation}\label{eq:Schip}
n_\alpha(0)\, \bbarb{S}\,\barb{\chi}_\alpha^\prime(0) =
\bbarb{S}\, \bbarb{\rho}^\prime(0)\, \bbarb{A}\,\,
\barb{\chi}_\alpha(0) \,\, .
\end{equation}
Multiplying instead with $\bbarb{A}$ and remembering
$\bbarb{\rho}(0) = \bbarb{A}\, \bbarb{\rho}(0)\,
\bbarb{A}$, from which follows $\bbarb{A}\, \bbarb{\rho}
(0) = \bbarb{\rho}(0)\,\bbarb{A}$, and using also
$\bbarb{A}\, \barb{\chi}_\alpha(0) = \barb{\chi}_\alpha
(0)$, one obtains
\begin{equation}\label{eq:Achip1}
\left[ n_\alpha(0) - \bbarb{\rho}(0) \right] \bbarb{A}\,
\barb{\chi}_\alpha^\prime(0) = \left[ \bbarb{A}\, 
\bbarb{\rho}^\prime(0)\, \bbarb{A} - n_\alpha^\prime(0)
\right] \barb{\chi}_\alpha(0) \,\, .
\end{equation}
Lastly, multiplication of (\ref{eq:Ot1term}) with
$\barb{\chi}_\alpha^\dagger(0)$ from the left, using the
zero-mode property of $\barb{\chi}_\alpha(0)$
[Eq.~(\ref{eq:Ot0term})], yields
\begin{equation}\label{eq:n0p}
n_\alpha^\prime(0) = \barb{\chi}_\alpha^\dagger(0)\,
\bbarb{\rho}^\prime(0)\, \barb{\chi}_\alpha(0) \quad .
\end{equation}
Thus we obtain implicit determining relations for
$\bbarb{S}\,\barb{\chi}_\alpha^\prime(0)$, $\bbarb{A}\,
\barb{\chi}_\alpha^\prime(0)$ and $n_\alpha^\prime(0)$
in terms of the eigenvector $\barb{\chi}_\alpha(0)$ and
eigenvalue $n_\alpha(0)>0$ of the 2bRDM.

Rearranging Eq.~(\ref{eq:Schip}) straightforwardly yields
\begin{equation}\label{eq:transform0}
\bbarb{S}\,\barb{\chi}_\alpha^\prime(0) = -\frac{i}{\hbar
\, n_\alpha(0)}\, \bbarb{\tau} \, \barb{\chi}_\alpha(0)
\end{equation}
in terms of the universal transformer from
Eq.~(\ref{eq:TauZerothOrder}) in the main text.
Equation~(\ref{eq:transform0}) is indeed the $t\to 0$
limit of the more general relation
(\ref{eq:TauDefImpl}). The explicit expression for
$\bbarb{A}\, \barb{\chi}_\alpha^\prime(0)$ is found from
Eq.~(\ref{eq:Achip1}), but we omit this here.

We continue by analyzing Eq.~(\ref{eq:Ot2term}).
Multiplying from the left with $\bbarb{S}$ and using the
relations $\bbarb{S}\, \bbarb{\rho}(0) = \bbarb{0}$,
$\bbarb{S}\, \barb{\chi}_\alpha(0) = \barb{0}$ and
$\bbarb{S}\,\bbarb{\rho}^\prime(0)\,\bbarb{S} =
\bbarb{0}$, we obtain
\begin{align}\label{eq:symmDPrel}
n_\alpha(0)\,\bbarb{S}\,\barb{\chi}_\alpha^{\prime\prime}
(0) &= \bbarb{S}\, \bbarb{\rho}^{\prime\prime}(0)\,
\bbarb{A} \, \barb{\chi}_\alpha(0) + 2\, \bbarb{S}\,
\bbarb{\rho}^{\prime}(0)\, \bbarb{A}\,
\barb{\chi}_\alpha^{\prime}(0) \nonumber \\
& \hspace{2.5cm} - 2\, n_\alpha^\prime(0)\, \bbarb{S}\,
\barb{\chi}_\alpha^{\prime}(0) \,\, .
\end{align}
Multiplying instead  with $\bbarb{A}$, we find
\begin{align}\label{eq:AchiDPimpl}
& \left[ n_\alpha(0) - \bbarb{\rho}(0) \right] \bbarb{A}
\, \barb{\chi}_\alpha^{\prime\prime}(0) = \left[
\bbarb{A}\,\bbarb{\rho}^{\prime\prime}(0)\, \bbarb{A} -
n_\alpha^{\prime\prime}(0) \right] \barb{\chi}_\alpha(0)
\nonumber \\ & \hspace{1em} + 2\, \bbarb{A}\,
\bbarb{\rho}^\prime(0)\, \barb{\chi}_\alpha^\prime(0) -
2\, n_\alpha^\prime(0)\, \bbarb{A}\,
\barb{\chi}_\alpha^\prime(0) \quad .
\end{align}
Finally, multiplying (\ref{eq:Ot2term}) with
$\barb{\chi}_\alpha^\dagger(0)$ from the left, using
also Eq.~(\ref{eq:Ot0term}), we find
\begin{align}\label{eq:n0pp}
n_\alpha^{\prime\prime}(0) &= \barb{\chi}_\alpha^\dagger
(0)\, \bbarb{\rho}^{\prime\prime}(0)\,\barb{\chi}_\alpha
(0) \nonumber \\ & \hspace{1cm} + 2\,
\barb{\chi}_\alpha^\dagger(0) \left[
\bbarb{\rho}^{\prime}(0) - n_\alpha^\prime(0) \right]
\barb{\chi}_\alpha^\prime(0)\quad .
\end{align}
Thus we have obtained implicit relations for all relevant
quantities to order $t^2$. In particular,
Eq.~(\ref{eq:symmDPrel}) yields
\begin{align}\label{eq:symmDPfin}
\bbarb{S}\,\barb{\chi}_\alpha^{\prime\prime}(0) &=
\frac{1}{n_\alpha(0)}\,\Big[ \bbarb{S}\,
\bbarb{\rho}^{\prime\prime}(0)\, \bbarb{A} \,
\barb{\chi}_\alpha(0) + 2\, \bbarb{S}\,
\bbarb{\rho}^{\prime}(0)\, \bbarb{A}\,
\barb{\chi}_\alpha^{\prime}(0) \nonumber \\
& \hspace{2.5cm} - 2\, n_\alpha^\prime(0)\, \bbarb{S}\,
\barb{\chi}_\alpha^{\prime}(0) \Big] \,\, ,
\end{align}
consistent with the general relation
(\ref{eq:TauDefImpl}). 

The perturbative scheme with small parameter $t$ for
obtaining eigenvectors and eigenvalues of the T2bCM
that have $n_\alpha(0)>0$ sketched above can be
straighforwardly extended to higher orders. The general
structure of the obtained perturbative expressions is
analogous to those found for the eigenvalues and
eigenvectors of the Hamiltonian in time-independent
nondegenerate perturbation theory of quantum
mechanics~\cite{Sakurai2011}. Results found for the
symmetric contribution to the eigenvector accord with
its general $t$ dependence discussed in
Appendix~\ref{app:genTrans}.

\subsection{Eigenvectors with vanishing eigenvalues at
\texorpdfstring{$\bm{t=0}$}{$t=0$}}\label{app:symmPert}

We now consider eigenvectors of the T2bCM whose
eigenvalue vanishes in the $t=0$ limit; $n_\alpha(0)=0$.
Due to the positive-semidefiniteness of T2bCM
eigenvalues, i.e., $n_\alpha(t)\ge 0$, $n_\alpha^\prime
(0)=0$ must also hold in this case, and
Eq.~(\ref{eq:eigenVexp}) specializes to
\begin{equation}
n_\alpha(t) = \frac{t^2}{2}\, n_\alpha^{\prime\prime}(0)
+ \mathcal{O}(t^3) \quad .
\end{equation}

The fact that $\bbarb{S}\,\bbarb{\rho}(t)\, \bbarb{S} =
t^2 \bbarb{S}\,\bbarb{\rho}^{\prime\prime}(0)\,\bbarb{S}
\, /2 + \mathcal{O}(t^3)$ implies that eigenvectors that
are entirely in the fully symmetric sector of the T2bCM
must have vanishing eigenvalues in the $t=0$ limit.
However, eigenvectors with $n_\alpha(0)=0$ may also
exist in the antisymmetric sector, which would be
signaled by having zero modes in the 2bRDM $\bbarb{\rho}
(0)$. We now present a careful treatment of the most
general case where there exists such a degeneracy
between symmetric and antisymmetric subspaces.

We start by defining projectors onto the degenerate and
nondegenerate subspaces. Specifically, we introduce
\begin{equation}
\Cproj = \sum^{\prime}_\alpha \barb{\chi}_\alpha(0)\,
\barb{\chi}_\alpha^\dagger(0) \quad ,
\end{equation}
where $\sum^{\prime}_\alpha$ indicates the restricted sum
over states having $n_\alpha(0)>0$. Its complement is
\begin{equation}
\Dproj \equiv \bbarb{1} - \Cproj = \sum^{\prime
\prime}_\beta \barb{\chi}_\beta(0)\,
\barb{\chi}_\beta^\dagger(0) \quad ,
\end{equation}
with $\sum^{\prime\prime}_\beta$ being the restricted sum
over states having $n_\beta(0)=0$. Eigenvectors from the
degenerate subspace satisfy $\barb{\chi}_\alpha(0)=\Dproj
\,\barb{\chi}_\alpha(0)$. Also, by construction, we have
$\bbarb{\rho}(0) \equiv \Cproj\, \bbarb{\rho}(0)\,
\Cproj$ and
\begin{equation}\label{eq:ProjRho0}
\Dproj\, \bbarb{\rho}(0) = \bbarb{\rho}(0)\,\Dproj =
\bbarb{0} \quad .
\end{equation}
We now analyze Eqs.~(\ref{eq:Ot012terms}) using the
projectors $\Dproj$ and $\Cproj$, having set
$n_\alpha(0)=0$ and $n_\alpha^\prime(0)=0$.

Multiplying Eq.~(\ref{eq:Ot1term}) from the left by
$\Dproj$, using also the relations (\ref{eq:ProjRho0})
and remembering that both $n_\alpha(0)$ and
$n_\alpha^\prime(0)$ vanish, we find $\Dproj\,
\bbarb{\rho}^\prime(0)\, \barb{\chi}_\alpha(0)=
\bbarb{0}$ for any eigenvector from the degenerate
subspace, implying
\begin{equation}\label{eq:PrhoprimeP}
\Dproj\,\bbarb{\rho}^\prime(0)\,\Dproj = \bbarb{0}\quad .
\end{equation}
Muliplying (\ref{eq:Ot1term}) with $\Cproj$ instead and
rearranging yields
\begin{equation}\label{eq:detQchi}
\Cproj\, \barb{\chi}_\alpha^\prime(0) = -\Cproj\,
\big[\bbarb{\rho}(0)\big]^{-1}\Cproj\,
\bbarb{\rho}^\prime(0)\,\Dproj\,\barb{\chi}_\alpha(0)
\quad ,
\end{equation}
where the inverse on the r.h.s.\ is well defined because
$\Cproj$ projects onto the subspace where $\bbarb{\rho}
(0)$ has no zero modes and is thus invertible.

Inserting $n_\alpha(0)=0$ and $n_\alpha^\prime(0)=0$ into
Eq.~(\ref{eq:Ot2term}), multiplying from the left with
$\Dproj$ and using the relation (\ref{eq:PrhoprimeP}), we
find
\begin{equation}
n_\alpha^{\prime\prime}(0)\, \barb{\chi}_\alpha(0) =
\Dproj\,\bbarb{\rho}^{\prime\prime}(0)\,\Dproj \,
\barb{\chi}_\alpha(0) + 2\, \Dproj\,
\bbarb{\rho}^{\prime}(0)\, \Cproj\,
\barb{\chi}_\alpha^\prime(0) \,\, .
\end{equation}
Utilising the explicit expression for $\Cproj\,
\barb{\chi}_\alpha^\prime(0)$ given in
Eq.~(\ref{eq:detQchi}) yields the eigenvalue equation
(\ref{eq:genEigenZero}) whose solution determines all
the $\barb{\chi}_\alpha(0)$ and their associated
$n_\alpha^{\prime\prime}(0)$.

In principle, the eigenvector $\chi_\alpha(t)$ can be
constructed order-by-order in $t$ by continuing the
perturbative treatment. However, for our purposes,
knowledge of $\barb{\chi}_\alpha(0)$ suffices. In
particular, if a single eigenvalue $n_0^{\prime\prime}
(0)$ among those obtained via
Eq.~(\ref{eq:genEigenZero}) turns out to be macroscopic,
then pairing order exists in the system at finite $t$,
even though no order is present for $t=0$. Most
generally, the macroscopic eigenvector $\barb{\chi}_0
(0)$ can be an arbitrary superposition of symmetric and
antisymmetric parts. The system exhibits
symmetric-pairing (i.e., odd-frequency) order when
$\bbarb{S}\,\barb{\chi}_0(0)$ is finite.

Further simplifications arise for eigenvectors that are
fully symmetric in the $t=0$ limit. Left-multiplying
Eq.~(\ref{eq:genEigenZero}) by $\bbarb{S}$, assuming
$\barb{\chi}_\beta(0) = \bbarb{S}\,\barb{\chi}_\beta(0)$,
and using $\bbarb{S}\,\Dproj = \Dproj\,\bbarb{S} =
\bbarb{S}$ gives
\begin{align}
\left\{ \bbarb{S}\,\bbarb{\rho}^{\prime\prime}(0)\,
\bbarb{S} - \frac{2}{\hbar^2}\,\bbarb{\tau}\,\Cproj \big[
\bbarb{\rho}(0) \big]^{-1} \Cproj\,\bbarb{\tau}^\dag
\right\} \barb{\chi}_\beta(0) = n_\beta^{\prime\prime}(0)
\, \barb{\chi}_\beta(0) \, ,
\end{align}
which is equivalent to the eigenvalue equation for the
generator matrix $\bbarb{\gamma}$
[Eq.~(\ref{eq:GammaEigEq}) from the main text].

\section{Relating eigenvalues of the generator matrix to
eigenvalues of the T2bCM}\label{app:genEvT2bCMEv}

A fundamental connection can be established between the
eigenvalues of the generator matrix $\bbarb{\gamma}$ and
the eigenvalues $n''_\beta(0)$ from
Eq.~(\ref{eq:genEigenZero}). Considering the spectral
decomposition
\begin{equation}
\bbarb{\gamma} = \sum_\nu g_\nu\, \barb{\lambda}_\nu\,
\barb{\lambda}_\nu^\dagger
\end{equation}
in conjunction with the expression given in
Eq.~(\ref{eq:genSchiSmallT}), one obtains the sum rule
\begin{equation}\label{eq:sumRule}
g_\nu = \frac{\hbar^2}{2} \sum_\beta^{\prime\prime}
n^{\prime\prime}_\beta(0)\, \big|
\barb{\lambda}_\nu^\dagger\, \barb{\chi}_\beta(0)\big|^2
\quad .
\end{equation}
For the special case where an eigenvector
$\barb{\lambda}_\nu$ coincides with one of the
$\barb{\chi}_\beta(0)$ and is therefore orthogonal to
all the others, Eq.~(\ref{eq:sumRule}) yields $g_\nu =
\hbar^2 n^{\prime\prime}_\beta(0)/2$, consistent with
Eq.~(\ref{eq:GammaEigEq}) from the main text. More
generally, specifics of the relationship between the
$g_\nu$ and the $n^{\prime\prime}_\beta(0)$ are encoded
by the overlap matrix $\big|\barb{\lambda}_\nu^\dagger\,
\barb{\chi}_\beta(0)\big|^2$.

Here we will prove that a macroscopic eigenvalue of
$\bbarb{\gamma}(t)$ implies the existence of a
macroscopic eigenvalue $n_\beta(t)\sim N\, \mathcal{O}
(t^2)$ of the T2bCM, under the further assumption that
the corresponding eigenvector has a finite overlap with
some eigenvector  of the T2bCM. More specifically, the
assumption is that the overlap of the respective
eigenvectors remains finite for large $N$. As the image
of the generator matrix coincides with the entire
symmetric subspace of the T2bCM, this assumption is
reasonable. 

We start by defining more precisely what we mean by a
macroscopic eigenvalue of $\bbarb{\gamma}(t)$. Let us
assume that there is a systematic way to change the
number $N$ of fermions in the system. Then an eigenvalue
$g_\nu(t)$ of $\bbarb{\gamma}(t)$ is macroscopic if
\begin{align} \label{eq:macroEig}
\frac{g_\nu(t)}{N} \ge c > 0 \quad ,
\end{align}
for all $N>N_c$ for some constants $c$ and $N_c$. 

Now let $g_0(t)$ be a macroscopic eigenvalue of
$\bbarb{\gamma}(t)$, with constants $c$ and $N_c$ as
explained above, satisfying
\begin{align} 
\bbarb{\gamma}(t)\, \barb{\lambda}_0(t) = g_0(t) \,
\barb{\lambda}_0(t) \quad .
\end{align}
Let $\barb{\chi}_\beta(t)$ be an eigenvector of the
T2bCM with a vanishing eigenvalue $n_\beta(0)$ at $t=0$. 
Let us further assume that the overlap of
$\barb{\chi}_\beta(t)$ with $\barb{\lambda}_0(t)$ is
bounded from below,
\begin{align}
\left|\barb{\lambda}^\dag_0(t)\, \barb{\chi}_\beta(t)
\right|^2 \ge \delta > 0 \quad ,
\end{align}
for all $N>N_\delta$ for some constants $\delta$ and
$N_\delta$.

We can now derive an inequality for the eigenvalue
$n_\beta(t)$ of the T2bCM:
\begin{align}
n_\beta(t) &= \barb{\chi}^\dag_\beta(t)\,\bbarb{\rho}(t)
\,\barb{\chi}_\beta(t) \nonumber \\
&\ge \barb{\chi}^\dag_\beta(t)\, \bbarb{S}\, \bbarb{\rho}
(t)\, \bbarb{S}\, \barb{\chi}_\beta(t) \nonumber \\
&\ge \frac{t^2}{\hbar^2}\, \barb{\chi}^\dag_\beta(t)\,
\bbarb{\gamma}(t)\, \barb{\chi}_\beta(t) \nonumber \\
&\ge g_0 \, \frac{t^2}{\hbar^2} \left|
\barb{\lambda}^\dag_0(t)\, \barb{\chi}_\beta(t)\right|^2
\nonumber \\
&\ge g_0\, \frac{t^2}{\hbar^2}\, \delta \nonumber \\
&\ge N \, t^2\, \frac{c\, \delta}{\hbar^2} \quad ,
\end{align}
where the last two lines only hold for sufficiently
large $N$. We have made use of the properties of
$\bbarb{\rho}(t)$ and $\bbarb{\gamma}(t)$ being
positive-semidefinite, as well as the fact that
$\bbarb{S}$ is a projector. If we have to weaken the
assumptions to hold only for $t\to 0$, then, by
continuity, the last line will still hold for a small
interval around $t=0$. 

\section{Relating order-parameter definitions of
number-conserving and nonconserving formalisms.}
\label{app:relateOPs}

Particle-nonconserving descriptions of pairing focus on
the anomalous pair-correlation function $\barb{F}(t_1,
t_2)$ defined via Eq.~(\ref{eq:AnomCorr}). This quantity
emerges from postulating~\cite{Gorkov1958,Cohen1965} the
factorization of the time-ordered two-particle
correlation function,
\begin{align}\label{eq:GorkovFact}
\Langle T\, c^\dagger_{\vek{i}}(t_1)\,c^\dagger_{\vek{j}}
(t_2)\, c_{\vek{l}}(t_2)\,c_{\vek{k}}(t_1) \Rangle \to
F_{\vek{i}\vek{j}}(t_1,t_2)\, F^*_{\vek{k}\vek{l}}(t_1,
t_2)\,\, ,
\end{align}
in the presence of a pair condensate. In contrast, the
generalized Penrose-Onsager-type approach developed here
considers the approximate factorization of the T2bCM
$\bbarb{\rho}(t,0)\equiv \bbarb{\rho}(t)$ to leading
order in $N$ [Eq.~(\ref{eq:T2bCMfact})] in terms of the
order parameter $\barb{\phi}(t)$
[Eq.~(\ref{eq:purelyASrho0})].

Although $\barb{F}(t,0)$ and $\barb{\phi}(t)$ are very
different quantities, they can be linked
conceptually~\cite{Leggett2006}
\begin{align}
\barb{F}(t,0)\, \longleftrightarrow\, \barb{\phi}(t)
\quad .
\end{align}
The even-frequency and odd-frequency parts of $\barb{F}
(t,0)$ correspond to its antisymmetric and symmetric
contributions \cite{Linder2019,Berezinskii1974},
\begin{subequations}
\begin{align}
\barb{F}^\mathrm{(e)}(t,0) \equiv \bbarb{A}\,\barb{F}
(t,0) \, \longleftrightarrow\, \barb{\phi}^\mathrm{(a)}
(t) \,\, , \\
\barb{F}^\mathrm{(o)}(t,0) \equiv \bbarb{S}\,\barb{F}
(t,0) \, \longleftrightarrow\, \barb{\phi}^\mathrm{(s)}
(t) \,\, .
\end{align}
\end{subequations}
In situations where only leading-order terms in the
limit $t\to 0$ are considered relevant, the quantities
\begin{subequations}
\begin{eqnarray}
\Delta^\mathrm{(e)} \equiv \barb{F}(0,0) &\,
\longleftrightarrow\, & \barb{\phi}^\mathrm{(a)}(0)
\,\, , \\
\Delta^\mathrm{(o)} \equiv \left. \partial_t\,\barb{F}
(t,0)\right|_{t=0} &\, \longleftrightarrow\, & \lim_{t
\to 0}\, \barb{\phi}^\mathrm{(s)}(t) /t
\end{eqnarray}
\end{subequations}
are sometimes used as the order
parameters~\cite{Abrahams1995,Dahal2009,Linder2019}.

\section{Interaction-related contribution to
transformer-induced symmetric-pairing order}
\label{app:UtauTriv}

We consider the specific case of a Zeeman-spin-split Fermi
superfluid discussed in Sec.~\ref{sec:ExamplesTransf}.
Using Eq.~(\ref{eq:TauSpInt}), the symmetric-pairing
order parameter~(\ref{eq:symmOPtr}) for this model
system can be decomposed as
\begin{align}
\barb{\phi}^\mathrm{(s)}(t) =
\barb{\phi}^{(\mathrm{s,sp})}(t) +
\barb{\phi}^{(\mathrm{s,int})}(t) \quad ,
\end{align}
where the leading-order small-$t$ expression of
$\barb{\phi}^{(\mathrm{s,sp})}(t)$ is given in
Eq.~(\ref{eq:SPhiZeem}), and 
\begin{align}
\barb{\phi}^{(\mathrm{s,int})}(t) = \frac{t}{\hbar}\,
\frac{1}{\sqrt{n_0(0)}}\,\,\bbarb{\tau}^{(\mathrm{int})}
\, \bbarb{A}\,\barb{\chi}_0(0)
\end{align}
to leading order in small $t$. Taking the explicit form
of the interaction part of the transformer
$\bbarb{\tau}^{(\mathrm{int})}$ from
Eq.~(\ref{eq:UtoTau}) and utilising also
Eq.~(\ref{eq:asymmEV}), we obtain
\begin{widetext}
\begin{equation}\label{eq:SPhiIntInterm}
\phi^{(\mathrm{s,int})}_{\,\vek{q}_i\sigma_i\, \vek{q}_j
\sigma_j}(t) = -U\, \frac{t}{\hbar}\,\frac{1}{\sqrt{n_0
(0)}}\,\, \Big\langle \sum_\vek{q} c^\dag_{\vek{q}
\uparrow}\, c^\dag_{-\vek{q}\downarrow}\left(\varsigma_j
\, c^\dag_{\vek{q}_i\sigma_i}\, c_{-\vek{q}_j
\bar{\sigma}_j} + \varsigma_i\, c^\dag_{\vek{q}_j
\sigma_j}\,c_{-\vek{q}_i \bar{\sigma}_i} \right)
\sum_{\vek{q'}}\chi_{0,\vek{q'}}\, c_{-\vek{q'}
\downarrow}\, c_{\vek{q'}\uparrow} \Big\rangle \,\,\, .
\end{equation}
\end{widetext}
Here we used again our compact notation where
$\bar{\sigma}$ denotes the opposite of $\sigma$; i.e.,
$\bar{\sigma} =\,\, \downarrow (\uparrow)$ if $\sigma =\,
\,\uparrow (\downarrow)$, and $\varsigma$ takes the
values $+1\, (-1)$ when $\sigma =\,\,\uparrow
(\downarrow)$.

To gain further insight into the general form of
$\barb{\phi}^{(\mathrm{s,int})}$, we adopt the
Yang-model description of a Fermi
superfluid~\cite{Yang1962}, where the ground state is
the pair-condensate state
\begin{equation}
\ket{\Psi_N} = \mathcal{N}_N \left( B^\dagger
\right)^\frac{N}{2}\, \ket{\mathrm{vac}}\,\, ,
\end{equation}
involving the pair-creation operator
\begin{equation}
B^\dagger = \frac{1}{\sqrt{m}} \sum_\vek{q}
c^\dag_{\vek{q}\uparrow}\,c^\dag_{-\vek{q}\downarrow}
\,\, .
\end{equation}
Here $m\gg N$ indicates the number of single-particle
modes (excluding spin), $\ket{\mathrm{vac}}$ is the
vacuum state, and the normalization factor
$\mathcal{N}_N$ is in principle known but does not need
to be specified here. For our purposes, we only need to
employ the relation
\begin{equation}\label{eq:Brelation}
B \ket{\Psi_N} = \left[ \frac{N}{2} - \frac{N(N-2)}{4 m}
\right]^\frac{1}{2} \ket{\Psi_{N-2}}
\end{equation}
and the known form of entries in the macroscopic
eigenvector of the 2bRDM [see Eq.~(\ref{eq:asymmEV})],
\begin{equation}\label{eq:asymmEVyang}
\chi_{0,\vek{q}} = \frac{1}{\sqrt{2m}}\,\, .
\end{equation}

With the input of Eqs.~(\ref{eq:Brelation}) and
(\ref{eq:asymmEVyang}), assuming also that the
expectation value on the r.h.s.\ of
Eq.~(\ref{eq:SPhiIntInterm}) is calculated in the Yang
state $\ket{\Psi_N}$, the interaction contribution to the
symmetric-pairing order parameter becomes
\begin{widetext}
\begin{subequations}
\begin{align}\label{eq:SPhiInt}
\phi^{(\mathrm{s,int})}_{\,\vek{q}_i\sigma_i\, \vek{q}_j
\sigma_j}(t) &= -U\, \frac{t}{\hbar}\,\frac{1}{\sqrt{n_0
(0)}}\,\, \sqrt{2m}\, \bra{\Psi_N} B^\dagger \big(
\varsigma_j\, c^\dag_{\vek{q}_i\sigma_i}\, c_{-\vek{q}_j
\bar{\sigma}_j} + \varsigma_i\, c^\dag_{\vek{q}_j
\sigma_j}\,c_{-\vek{q}_i\bar{\sigma}_i} \big) B
\ket{\Psi_N} \,\, , \\ \label{eq:intTrans}
&= -m\, U\,\, \frac{t}{\hbar}\,\, \frac{N}{\sqrt{n_0(0)}}
\, \left[ 1 - \frac{N-2}{2m}\right]\,\frac{1}{\sqrt{2m}}
\, \bra{\Psi_{N-2}}\big(\varsigma_j\, c^\dag_{\vek{q}_i
\sigma_i}\, c_{-\vek{q}_j \bar{\sigma}_j} + \varsigma_i
\, c^\dag_{\vek{q}_j \sigma_j}\,c_{-\vek{q}_i
\bar{\sigma}_i} \big) \ket{\Psi_{N-2}} \,\, . 
\end{align}
\end{subequations}
On the r.h.s.\ of Eq.~(\ref{eq:intTrans}), normalization
factors have been distributed such that the result
emerging in the limits of $m\to\infty$ and large $N\ll
m$ is readily apparent. In particular, with $n_0(0)\sim
\mathcal{O}(N)$, it appears that, as far as the scaling
as a function of $N$ is concerned,
$\barb{\phi}^{(\mathrm{s,int})}(t)$ could be a relevant
contribution to the symmetric-pairing order parameter.

The vector entries of $\barb{\phi}^{(\mathrm{s,int})}
(t)$ given in Eq.~(\ref{eq:intTrans}) are related to a
combination of entries from the single-particle reduced
density matrix for the $N-2$-particle Yang state. This
again illustrates the general importance of
single-particle physics for facilitating the
transformation of antisymmetric pairing order into
symmetric pairing order. Within the Yang-model
description, exact results for the single-particle
reduced density matrix yield
\begin{align}
\bra{\Psi_{N-2}} c^\dag_{\vek{q}_i \sigma_i}\,
c_{-\vek{q}_j \bar{\sigma}_j} \ket{\Psi_{N-2}} =
\bra{\Psi_{N-2}} c^\dag_{\vek{q}_j \sigma_j}\,
c_{-\vek{q}_i \bar{\sigma}_i} \ket{\Psi_{N-2}} =
\frac{N-2}{2 m}\,\, \delta_{\vek{q}_j, -\vek{q}_i}\,
\delta_{\sigma_j, \bar{\sigma}_i} \,\, ,
\end{align}
and we find
\begin{equation}\label{eq:vanishIntSymm}
\phi^{(\mathrm{s,int})}_{\,\vek{q}_i\sigma_i\, \vek{q}_j
\sigma_j}(t) \propto -\varsigma_i\bra{\Psi_{N-2}} \big(
c^\dag_{\vek{q}_i \sigma_i}\, c_{\vek{q}_i \sigma_i} -
c^\dag_{-\vek{q}_i \bar{\sigma}_i}\,c_{-\vek{q}_i
\bar{\sigma}_i} \big) \ket{\Psi_{N-2}} \,\,
\delta_{\vek{q}_j, -\vek{q}_i}\, \delta_{\sigma_j,
\bar{\sigma}_i} = 0\,\, .
\end{equation}
Thus, within the Yang-model description, the interaction
part of the transformer does not cause symmetric-pairing
order, and the latter originates entirely from the
single-particle portion of the transformer that gives
rise to the order parameter from Eq.~(\ref{eq:SPhiZeem}).

The fundamental reason why the interaction-related part
of the transformer yields no symmetric-pairing order can
be gleaned from inspecting the correlation function
appearing on the r.h.s.\ of
Eq.~(\ref{eq:vanishIntSymm}). The two terms being
subtracted are related via the time-reversal operation,
which inverts both the spin and orbital momentum. This
observation leads us to surmise that the vanishing of
$\barb{\phi}^{(\mathrm{s,int})}(t)$ holds more generally
on symmetry grounds, even beyond the strict limits of
applicability for the Yang model.

\section{Generated symmetric-pairing order in a system
of itinerant electrons coupled to magnons}

\subsection{Form of the transformer and generator
matrices}\label{app:elMagTauGen}

Using the definition
Eq.~(\ref{eq:TransMat}), with $H$ given by
Eq.~(\ref{eq:HfHbHffb}), we find the expression
\begin{align}\label{eq:Tau1DLatt}
&\tau_{r_i\sigma_i\, r_j\sigma_j\, ,\, r_k\sigma_k\, r_l
\sigma_l} = \nonumber \\[0.2cm] & \hspace{1cm}
\frac{K}{2}\,\Big[ \, \rho_{r_i+1\,\sigma_i\, r_j
\sigma_j\, ,\, r_k\sigma_k\, r_l\sigma_l}(0) +
\rho_{r_i-1\,\sigma_i\, r_j\sigma_j\, ,\, r_k\sigma_k\,
r_l\sigma_l}(0) + \rho_{r_j+1\,\sigma_j\, r_i\sigma_i\,
,\, r_k\sigma_k\, r_l\sigma_l}(0) + \rho_{r_j-1\,\sigma_j
\, r_i\sigma_i\, ,\, r_k\sigma_k\, r_l\sigma_l}(0)\,
\Big] \nonumber \\ & \hspace{1.5cm} +\,
\frac{J}{2}\, \Big[\,\delta_{\sigma_i,\uparrow}\,\Langle
c^\dag_{r_j\sigma_j}\, c^\dag_{r_i\downarrow}\,
b^\dag_{r_i}\, c_{r_l\sigma_l}\, c_{r_k\sigma_k} \Rangle
+ \delta_{\sigma_i,\downarrow}\,\Langle c^\dag_{r_j
\sigma_j}\, c^\dag_{r_i\uparrow}\, b_{r_i}\, c_{r_l
\sigma_l}\, c_{r_k\sigma_k}\Rangle \nonumber \\
& \hspace{6.5cm} +\, \delta_{\sigma_j,\uparrow}\,\Langle
c^\dag_{r_i\sigma_i}\,c^\dag_{r_j\downarrow}\,
b^\dag_{r_j}\, c_{r_l\sigma_l}\,c_{r_k\sigma_k} \Rangle
+ \delta_{\sigma_j,\downarrow}\,\Langle c^\dag_{r_i
\sigma_i}\,c^\dag_{r_j\uparrow}\,b_{r_j}\,c_{r_l
\sigma_l}\, c_{r_k\sigma_k} \Rangle \Big]
\end{align}
for the transformer. The terms $\propto J$ in
Eq.~(\ref{eq:Tau1DLatt}) are expectation values
involving unbalanced boson creation and annihilation
operators $b^\dag_r$ and $b_r$. If the state defining the
expectation values in Eq.~(\ref{eq:Tau1DLatt}) is an
eigenstate of the boson number operator
$\hat{N}_\mathrm{b} = \sum_r b^\dag_r b_r$, then these
terms must vanish identically. 

Expectation values involving unbalanced boson operators
also arise in the calculation of the part $\propto
\bbarb{S}\,\bbarb{\rho}^{\prime\prime}(0)\,\bbarb{S}$ of
the generator, see the first term in
Eq.~(\ref{eq:gammaForm}), and they are neglected on the
same grounds in that context. The remaining terms
$\propto K^2$ arising due to the $\bbarb{S}\,
\bbarb{\rho}^{\prime\prime}(0)\,\bbarb{S}$ part are
cancelled by the quadratic-in-$\bbarb{\tau}$
contribution to the generator [see the second term in
Eq.~(\ref{eq:gammaForm})]. The result for the generator
depends only on terms $\propto J^2$, and can be
expressed as Eq.~(\ref{eq:Gamma1DLatt}) of the main
text. 

\subsection{Reduced density matrices for the composite
fermion-pair--magnon condensate}\label{app:compCondRDM}

To demonstrate the typical behaviour of reduced density
matrices corresponding to composite-condensate
systems~\cite{Dahal2009}, we consider a Yang-type state
for composite bosons,
\begin{subequations}\label{eq:PhiNBDefs}
\begin{align}\label{eq:PhiDef}
\ket{\Psi_N^\mathrm{(3b)}} &=
\mathcal{N}_N\,\big( B^\dag_\mathrm{fb}\big)^\frac{N}{2}
\ket{\mathrm{vac}} \quad , \\ \label{eq:BcompDef} 
B^\dag_\mathrm{fb} &= \frac{1}{\sqrt{m}}\,\sum_r
c^\dag_{r\uparrow} c^\dag_{r\downarrow} b^\dag_r\quad ,
\\ \label{eq:BrelDahal}
B_\mathrm{fb} \ket{\Psi_N^\mathrm{(3b)}} &= \left[
\frac{N}{2} - \frac{N(N-2)}{4 m} \right]^\frac{1}{2}
\ket{\Psi_{N-2}^\mathrm{(3b)}} \quad ,
\end{align}
\end{subequations}
where $m$ is the total number of one-dimensional lattice
sites. Equations~(\ref{eq:PhiNBDefs}) constitute a
particular example of a composite-boson condensate that
is an eigenstate of both the fermion and boson number
operators $\hat{N}$ and $\hat{N}_\mathrm{b}$. 

In the limit of a large lattice, i.e., for $m\gg N$, the
system's 2bRDM and the three-body reduced density matrix
defined by Eq.~(\ref{eq:3bRDM}), respectively, take the
form
\begin{subequations}\label{eq:compRDMs}
\begin{align}\label{eq:ApproxYD2bRDM} 
\bra{\Psi_N^\mathrm{(3b)}} c^\dag_{r_i\sigma_i}\,
c^\dag_{r_j\sigma_j}\,c_{r_l\sigma_l}\,c_{r_k\sigma_k}
\ket{\Psi_N^\mathrm{(3b)}} &\approx \frac{N}{2 m}\,\,
\delta_{\sigma_j,\bar{\sigma}_i}\, \delta_{r_i, r_j}\,
\delta_{r_l, r_k}\,\big(\delta_{r_i\sigma_i\, ,\, r_k
\sigma_k}\,\delta_{r_j\sigma_j\, ,\, r_l\sigma_l} -
\delta_{r_j\sigma_j\, ,\, r_k\sigma_k}\, \delta_{r_i
\sigma_i\,,\,r_l\sigma_l} \big) \,\,\, , \\
\bra{\Psi_N^\mathrm{(3b)}} c^\dag_{r_i\sigma_i}\,
c^\dag_{r_j\sigma_j}\, b^\dag_{r_o}\, b_{r_p} \, c_{r_l
\sigma_l}\, c_{r_k\sigma_k}\ket{\Psi_N^\mathrm{(3b)}}
&\approx \frac{N}{2m}\,\,\varsigma_i\,\varsigma_k\,\,
\delta_{r_i, r_j}\,\delta_{r_l, r_k}\, \delta_{r_i, r_o}
\,\delta_{r_p, r_k}\,\delta_{\sigma_j,\bar{\sigma}_i}\,
\delta_{\sigma_l,\bar{\sigma}_k} \nonumber \\[0.1cm]
&\equiv N \left[ \frac{1}{\sqrt{2 m}}\,\,\varsigma_i\,\,
\delta_{r_j, r_i}\,\delta_{r_o, r_i}\,\delta_{\sigma_j,
\bar{\sigma}_i}\right] \left[ \frac{1}{\sqrt{2m}}\,\,
\varsigma_k\,\, \delta_{r_l, r_k}\,\delta_{r_p, {r}_k}\,
\delta_{\sigma_l, \bar{\sigma}_k}\right]^* .
\label{eq:ApproxYD3bRDM}
\end{align}
\end{subequations}
\end{widetext}
We utilized again the notation where $\bar{\sigma}$ is
the opposite of $\sigma$; i.e., $\bar{\sigma} =\,\,
\downarrow (\uparrow)$ if $\sigma =\,\,\uparrow
(\downarrow)$, and $\varsigma$ takes the values $+1\,
(-1)$ when $\sigma =\,\,\uparrow (\downarrow)$. The
2bRDM (\ref{eq:ApproxYD2bRDM}) is a block-diagonal
matrix in two-particle index space with degenerate 
eigenvalues $0$ and $N/m \ll 1$ that are nonmacroscopic.
It can be verified that the vectors from three-body
index space appearing in Eq.~(\ref{eq:ApproxYD3bRDM})
are normalised. The rank-1 three-body reduced density
matrix thus has eigenvalue $N$, which is consequently
macroscopic, in the limit $m\gg N$. Identifying $1/
\sqrt{2m} = \chi_0^\mathrm{(3b)}$ yields
Eq.~(\ref{eq:3bRDMeigvec}) in the main text. 

\bibliography{Refs}

\end{document}